\documentclass[journal]{IEEEtran}
\usepackage{amsmath, amssymb}
\usepackage{graphicx}
\usepackage{cite}
\usepackage{url}
\usepackage{booktabs}  
\usepackage{placeins}
\usepackage{amsmath}
\usepackage{amssymb}
\usepackage{caption}
\usepackage[hidelinks]{hyperref}

\title{FlowSonic: Stable Zero-Shot Music Editing via High-Order Trajectory Integration}
\author{Ali Boudaghi,
        Hadi Zare%
\thanks{Corresponding author: Ali Boudaghi (email: ali.boudaghi@ut.ac.ir).}
\thanks{Code and implementation are publicly available at: \url{https://github.com/aliramsy/FlowSonic}.}}

\begin{document}

\maketitle

\begin{abstract}

Zero-shot text-guided editing of real-world music recordings requires balancing semantic modification with faithful preservation of the original musical structure. Although recent diffusion transformers trained with rectified flow have achieved remarkable success in text-to-music generation, extending them to edit existing recordings remains challenging because editing requires accurate deterministic inversion, reliable structural preservation, and numerically stable integration throughout the inversion and generation processes.

We present FlowSonic, a zero-shot music editing framework built upon a pretrained diffusion transformer trained with rectified flow. FlowSonic first deterministically inverts a real-world recording into the latent space and preserves its musical structure during editing by reusing cross-attention representations extracted during inversion. To improve the numerical reliability of inversion-based editing, we introduce a high-order ODE solver and systematically investigate how different numerical integration schemes influence trajectory stability, structural preservation, and semantic controllability.

Comprehensive experiments on timbre-transfer and genre-modification tasks demonstrate that FlowSonic consistently outperforms existing music editing methods across semantic alignment, harmonic preservation, structural consistency, and perceptual audio quality. We further provide geometric and empirical analyses showing how the proposed numerical integration strategy improves latent trajectory stability and leads to more reliable music editing.

\end{abstract}


\section{Introduction}

Diffusion models have emerged as one of the most successful generative modeling paradigms, learning to synthesize high-dimensional data by progressively transforming random noise into structured samples through a denoising process~\cite{ddpm,scorebased,edm,flowmatching}.

Following their success in image generation, diffusion models have rapidly transformed audio generation, enabling high-quality synthesis of speech, sound, and music~\cite{stablediffusion,imagen,audioldm2,forsgren2022riffusion,huang2023noise2music,Aufussion,musicldm,fluxplaysmusic}.

Modern text-to-music models can synthesize complete musical pieces from natural-language descriptions by leveraging diffusion models and transformer architectures~\cite{forsgren2022riffusion, liu2023audioldm, huang2023noise2music, Aufussion, accomontage, musicldm, lin2023content, musecoco, copet2024simple, melechovsky2023mustango, textcond}. Despite their impressive capabilities, these systems are primarily designed for generating music from scratch rather than editing existing recordings.

In practical music production, however, editing is often more important than generation. Common tasks include refining performances, changing instrumentation, or transforming an existing recording into a different musical style. For musicians, producers, and content creators, modifying an existing piece is frequently more useful than creating one from scratch.

Music editing is considerably more challenging than music generation because it must simultaneously satisfy two competing objectives: applying the desired modification while preserving the musical characteristics that should remain unchanged. This balance becomes particularly difficult for expressive, polyphonic, and multi-instrument recordings. Existing approaches address this problem either through supervised learning on paired ``before'' and ``after'' datasets~\cite{m2ugen, instructme, audit} or through zero-shot latent manipulation of diffusion models~\cite{musicmagus, zeta, transplayer}. However, these methods are often tailored to specific editing tasks, primarily operate on model-generated rather than real-world audio, and frequently rely on carefully engineered prompts to produce satisfactory edits~\cite{musicmagus, transplayer}. These limitations reduce their flexibility and practicality for real-world creative workflows. Meanwhile, recent studies have also demonstrated the potential of diffusion models for audio restoration tasks, including equalization and bandwidth extension~\cite{moliner2024diffusion_equalizer}.

Recently, rectified flow models~\cite{lee2024improvingtrainingrectifiedflows, liu2022flowstraightfastlearning} have emerged as an efficient alternative to conventional diffusion models by learning a more direct transport between noise and data distributions. Their formulation enables straighter sampling trajectories and efficient generation while demonstrating excellent scalability in large transformer-based generative architectures~\cite{tang2024hart, xie2024sana, yang2024cogvideox, peebles2023scalable}. In particular, \emph{FLUX that Plays Music}~\cite{fluxplaysmusic} demonstrates the effectiveness of rectified-flow transformers for high-quality text-to-music generation.

Despite these advances, rectified-flow models have been explored primarily for forward generation rather than editing existing recordings. Extending them to real-world music editing introduces a substantially more challenging problem: an input recording must first be deterministically inverted into the latent space before guided generation can perform the desired semantic modification. Unlike standard text-to-music synthesis, inversion-based editing consists of two coupled stages, deterministic inversion followed by conditional generation, both of which rely on numerically solving the underlying rectified-flow dynamics. Numerical inaccuracies introduced during either stage can accumulate throughout the editing process and ultimately degrade structural preservation and editing quality. This raises a fundamental question: \emph{how can rectified-flow trajectories be integrated to enable stable, high-fidelity music editing while faithfully preserving the structure of the original recording?}

\subsection{Our Approach}

To address these challenges, we propose \textbf{FlowSonic}, a zero-shot framework for editing real-world music recordings using pretrained rectified-flow models. Rather than relying on additional training, optimization, or paired editing data, FlowSonic reformulates music editing as a deterministic inversion-and-generation process in the latent space of a pretrained text-to-music model.

Starting from a deterministic inversion of the input recording, FlowSonic preserves the musical structure of the source audio by reusing cross-attention representations throughout text-guided generation. To improve numerical accuracy, we further investigate the use of high-order multi-step ODE solvers for rectified-flow music editing and adopt the third-order Adams--Bashforth (AB3) integrator as a generic numerical framework for the generation process. We identify an initialization inconsistency that arises when applying multi-step integration to inversion-based editing and address it by introducing \emph{Dynamic History Caching(DHC)}, which initializes the AB3 solver using velocity evaluations obtained during deterministic inversion. By eliminating conventional lower-order warm-start initialization, the proposed strategy stabilizes the generation trajectory, improves numerical consistency, and enhances both structural preservation and semantic controllability without modifying the pretrained model or requiring additional optimization.

\subsection{Contributions}

The main contributions of this work are summarized as follows:

\begin{itemize}

\item We propose \textbf{FlowSonic}, a zero-shot framework for editing real-world music recordings using pretrained rectified-flow models without fine-tuning, optimization, or paired editing data.

\item We introduce a high-order numerical integration framework for inversion-based rectified-flow music editing based on the third-order Adams--Bashforth (AB3) solver, together with DHC to eliminate lower-order warm-start initialization.

\item We systematically investigate the impact of numerical integration strategies on rectified-flow music editing. Through latent trajectory visualization, Mel-spectrogram analysis, and objective evaluation, we analyze how different integration schemes influence trajectory stability, semantic editing strength, structural preservation, and perceptual audio quality.

\item Extensive experiments on timbre transfer and genre transformation demonstrate that FlowSonic consistently outperforms existing zero-shot editing methods while preserving musical structure and achieving strong semantic alignment with target editing prompts.

\end{itemize}

The remainder of this paper is organized as follows. Section~\ref{sec:related_work} reviews existing approaches to text-to-music generation and music editing. Section~\ref{sec:overview} presents the proposed FlowSonic framework, including the overall editing pipeline, deterministic inversion, and the proposed DHC strategy for stable high-order integration. Section~\ref{sec:experiments} describes the experimental protocol and reports extensive quantitative and qualitative evaluations on timbre-transfer and genre-modification tasks, together with ablation studies that analyze the effect of the proposed numerical integration strategy. Finally, Section~\ref{sec:conclusion} summarizes the main findings and outlines future research directions.

\begin{figure*}[!t]
  \centering
  \includegraphics[width=\textwidth]{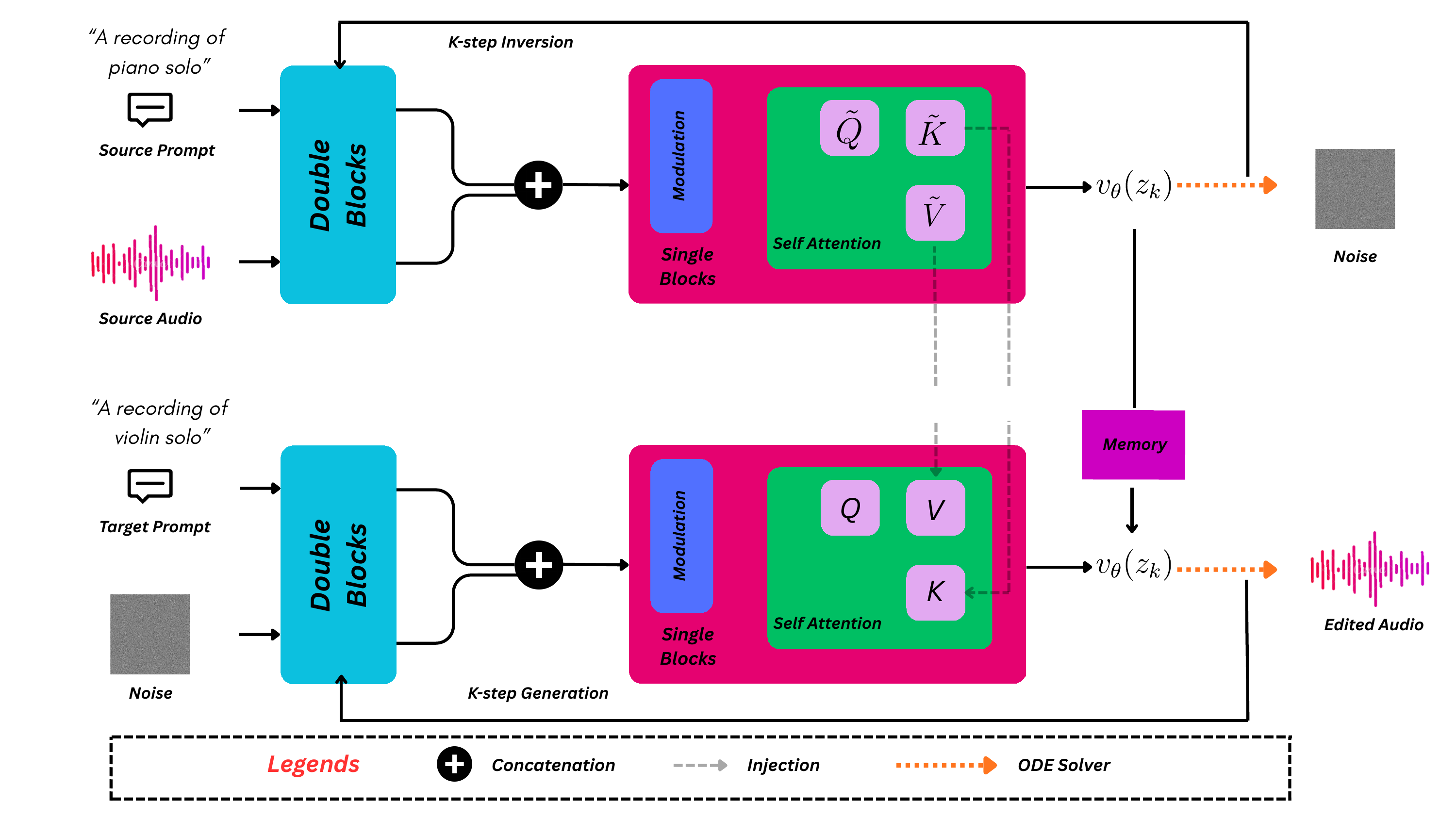}
  \caption{
Overview of the proposed FlowSonic framework for zero-shot text-guided music editing. During deterministic inversion (top), the source audio and its corresponding source prompt are mapped into the rectified-flow latent space while cross-attention representations are cached. The inversion process simultaneously records velocity evaluations that are later reused to initialize the high-order generation solver. During generation (bottom), the inverted latent is edited using a new target prompt, while the cached attention features are selectively injected to preserve musical structure. The stored velocity history initializes the generation-stage multi-step solver from the first editing step, avoiding lower-order warm-start initialization during generation and enabling stable trajectory regularization.}
  \label{fig:method}
\end{figure*}

\section{Related Work}
\label{sec:related_work}

\subsection{Text-to-Music Generation}

Recent years have witnessed substantial progress in text-to-music generation driven by advances in autoregressive, diffusion, and flow-based generative models. Early systems primarily relied on autoregressive language models adapted for audio synthesis~\cite{agostinelli2023musiclm, copet2024simple, musicgenstem, inspiremusic}. These models naturally capture long-range temporal dependencies and produce coherent sequential outputs, but autoregressive decoding is inherently sequential, making inference relatively slow while also allowing prediction errors to accumulate over long generations.

Diffusion-based approaches have since become a dominant paradigm for high-fidelity music synthesis. Models such as Riffusion~\cite{forsgren2022riffusion}, AudioLDM~\cite{audioldm2}, DiffRhythm~\cite{diffrythm}, Möusai~\cite{mousai}, and Tango~\cite{tango} generate realistic audio directly from textual descriptions through iterative denoising. Although diffusion models generally produce higher-quality outputs and are less susceptible to error accumulation than autoregressive approaches, their iterative sampling process often requires many inference steps, resulting in relatively high computational cost.

To combine the advantages of both paradigms, several hybrid architectures have been proposed. Approaches including Auffusion~\cite{Aufussion} and MagNet~\cite{magnet} integrate transformer-based sequence modeling with diffusion-based generation, improving both controllability and generation quality while reducing some of the limitations associated with each individual framework.

Beyond text conditioning, recent studies have incorporated additional control signals such as melody, harmony, and rhythmic structure to provide users with finer control over the generated music~\cite{ditto, musicontrolnet, cocomulla}. Despite these advances, the majority of existing methods remain focused on music generation rather than modification of existing recordings.

More recently, rectified-flow (RF) models have emerged as an efficient alternative to diffusion models for music generation~\cite{musflow, fluxplaysmusic} and editing~\cite{melodyflow}. By learning deterministic transport trajectories instead of stochastic denoising processes, RF models substantially reduce the sampling complexity while maintaining high synthesis quality. Their deterministic formulation also naturally supports inversion, making them particularly attractive for editing applications.

\subsection{Music Editing}

Compared with music generation, text-guided music editing remains a relatively underexplored problem. Existing approaches can generally be divided into two categories. The first relies on retraining or fine-tuning pretrained generative models for specific editing tasks~\cite{adapter}. Although these methods can achieve strong performance, they require task-specific supervision and repeated optimization whenever a new editing capability is introduced, limiting their scalability and practical applicability.

The second category performs editing in a zero-shot manner using pretrained generative models. Representative examples include \emph{MusicMagus}~\cite{musicmagus}, which manipulates latent representations within diffusion models to achieve text-guided editing without additional training. While promising, these methods are typically designed for model-generated content and often experience noticeable performance degradation when applied to arbitrary real-world recordings. Furthermore, they frequently depend on carefully engineered prompts to obtain reliable edits, reducing accessibility for non-expert users~\cite{musicmagus}.

Recent progress in rectified-flow models provides an opportunity to address several of these limitations. Their deterministic trajectories enable inversion of real audio into the latent space, allowing editing to be formulated as a controlled modification of the inversion trajectory. However, effectively combining deterministic inversion, structural feature preservation, and stable numerical integration remains an open challenge. In particular, high-order multi-step solvers offer improved trajectory accuracy but introduce initialization inconsistencies when coupled with inversion-based feature injection.

\section{Overview and Methodology}
\label{sec:overview}

Figure~\ref{fig:method} illustrates the overall pipeline of the proposed FlowSonic framework. Our goal is to edit an existing music recording according to a natural-language instruction while preserving the musical characteristics that should remain unchanged. Unlike conventional text-to-music generation, the editing process must first recover a latent representation of the input recording before performing controlled generation toward the target description.

The framework therefore consists of two consecutive stages. First, the input audio is deterministically inverted into the latent space of a pretrained diffusion transformer. During this process, internal cross-attention representations together with the numerical integration history are recorded for later reuse. In the second stage, the latent representation is guided toward the target prompt while the cached attention features are injected into the transformer to preserve the structural information of the original recording. To improve the numerical stability of the generation process, we propose \textit{FlowSonic}, which combines the proposed Seeded-AB3 solver with a DHC initialization mechanism to eliminate the lower-order warm-start phase normally required by multi-step integration. This combination enables accurate semantic editing while maintaining harmonic, melodic, and spectral consistency throughout the generated music.

\subsection{Rectified Flow}

Rectified flow~\cite{liu2022flowstraightfastlearning} formulates generation as a continuous-time transport process between a simple prior distribution and the target data distribution. Let $\pi_0$ denote the prior distribution (e.g., $\mathcal{N}(0,I)$) and $\pi_1$ denote the data distribution. Rather than learning a stochastic reverse diffusion process, rectified flow models a deterministic velocity field that transports samples from $\pi_0$ to $\pi_1$ along approximately straight trajectories.

Given paired samples $(z_0,z_1)$ drawn from a coupling of $\pi_0$ and $\pi_1$, the intermediate latent state at time $t\in[0,1]$ is defined by a linear interpolation

\begin{equation}
\label{eq:linear-interp}
z_t
=
\alpha_t z_0
+
\beta_t z_1,
\end{equation}

where $\alpha_0=1$, $\beta_0=0$, $\alpha_1=0$, and $\beta_1=1$. The canonical rectified-flow parameterization adopts
$\alpha_t=1-t$ and $\beta_t=t$, producing a straight-line trajectory between the source and target states.

Differentiating \eqref{eq:linear-interp} yields the target velocity along the transport path,

\begin{equation}
\label{eq:target-velocity}
\dot z_t
=
\dot\alpha_t z_0
+
\dot\beta_t z_1.
\end{equation}

Under the canonical linear schedule, this simplifies to the constant velocity
$\dot z_t = z_1-z_0$, reflecting the straight transport property that distinguishes rectified flow from conventional diffusion models.

To approximate this transport process, rectified flow learns a time-dependent velocity field $v_\theta(z,t)$ governed by the ordinary differential equation

\begin{equation}
\label{eq:rf-ode}
\frac{d z(t)}{dt}
=
v_\theta\bigl(z(t),t\bigr),
\end{equation}

where the objective is to transport samples initialized from $\pi_0$ toward the target distribution $\pi_1$. Training is performed through velocity matching by minimizing

\begin{equation}
\label{eq:velocity-loss}
\mathcal{L}(\theta)
=
\mathbb{E}_{(z_0,z_1),\,t}
\left[
\left\|
v_\theta(z_t,t)
-
\dot z_t
\right\|^2
\right],
\end{equation}

where $z_t$ and $\dot z_t$ are given by
\eqref{eq:linear-interp} and \eqref{eq:target-velocity}, respectively. For the canonical interpolation schedule, the network therefore learns the constant velocity $z_1-z_0$ along the transport path.

During inference, both generation and deterministic inversion are obtained by numerically integrating the learned velocity field in \eqref{eq:rf-ode}. A standard explicit Euler discretization over a partition
$0=t_0<t_1<\cdots<t_K=1$
updates the latent state as

\begin{equation}
\label{eq:euler-step}
z_{k+1}
=
z_k
+
(t_{k+1}-t_k)
v_\theta(z_k,t_k).
\end{equation}

Although first-order Euler integration is widely adopted because of its simplicity, the accumulated discretization error becomes increasingly important in inversion-based editing, where inaccuracies introduced during inversion directly influence the subsequent generation trajectory. This motivates the use of higher-order numerical integration during inversion and high-order multi-step integration during guided generation, where numerical errors have a greater impact on editing quality. The deterministic inversion process built upon this formulation is described in the following subsection.

\subsection{Deterministic Inversion}

Music editing requires mapping an observed recording back into the latent space from which it can be manipulated and regenerated. In diffusion-based generative models, this process is commonly referred to as \emph{inversion}. Since the forward denoising process is only approximated numerically, recovering the latent representation of a real sample is inherently challenging.

A widely adopted approach is \emph{DDIM inversion}~\cite{songscore, ddim}, which reconstructs the latent trajectory by sequentially reversing the deterministic DDIM sampling process. By repeatedly estimating the noise component at each timestep, DDIM inversion produces a latent representation that approximately reconstructs the input signal. However, because the trajectory is recovered through low-order numerical integration, discretization errors accumulate along the reverse path, causing the reconstructed latent to drift from the original signal. To improve inversion fidelity, several optimization-based and trajectory-alignment methods have been proposed~\cite{nti, stsl, elarabawy2022direct, negprompt}. Although these approaches often achieve better reconstruction quality, they typically require iterative optimization or additional correction procedures, increasing computational cost and limiting their practicality for efficient editing.

Rectified flow models provide a different perspective on inversion. Rather than learning stochastic diffusion trajectories, rectified flows learn a continuous velocity field that transports samples between the latent distribution and the data distribution through approximately straight probability paths. Under the continuous-time formulation, the reverse dynamics are theoretically deterministic. Given an observed sample $\mathbf{z}_1 \sim \pi_1$, inversion corresponds to integrating the learned velocity field backward from $t=1$ to $t=0$:

\begin{equation}
\label{eq:continuous-inversion}
\mathbf{z}(t-\Delta t)=
\mathbf{z}(t)-
\int_{t-\Delta t}^{t}
v_\theta\!\left(\mathbf{z}(\tau),\tau\right)
\,d\tau .
\end{equation}

Although this continuous formulation is conceptually simple, practical inversion still requires numerical discretization of the trajectory.

Research on inversion specifically for rectified-flow models remains relatively limited. Early methods such as \emph{RF-Prior}~\cite{yang2024text} formulate inversion through score-distillation optimization to recover the latent representation of an observed sample. While effective, these optimization procedures are computationally expensive and therefore less suitable for interactive editing. More recently,~\cite{rout2024rfinversion} proposed augmenting the generative model with an auxiliary vector field conditioned on the input signal to improve reconstruction accuracy. Although this approach enhances inversion quality, it does not directly address the numerical errors introduced by discretizing the original rectified-flow trajectory. Consequently, stable inversion and high-quality downstream editing remain challenging.

In practice, deterministic inversion is performed by integrating the rectified-flow ODE backward in time using the first-order Euler discretization introduced in Eq.~(\ref{eq:euler-step}), with the integration direction reversed from $t=1$ to $t=0$. The resulting latent representation serves as the starting point for the subsequent generation and editing process.

\subsection{High-Order Numerical Integration for Music Editing}
\label{sec:solver}

\begin{figure*}[!t]
\centering
\includegraphics[width=0.9\textwidth]{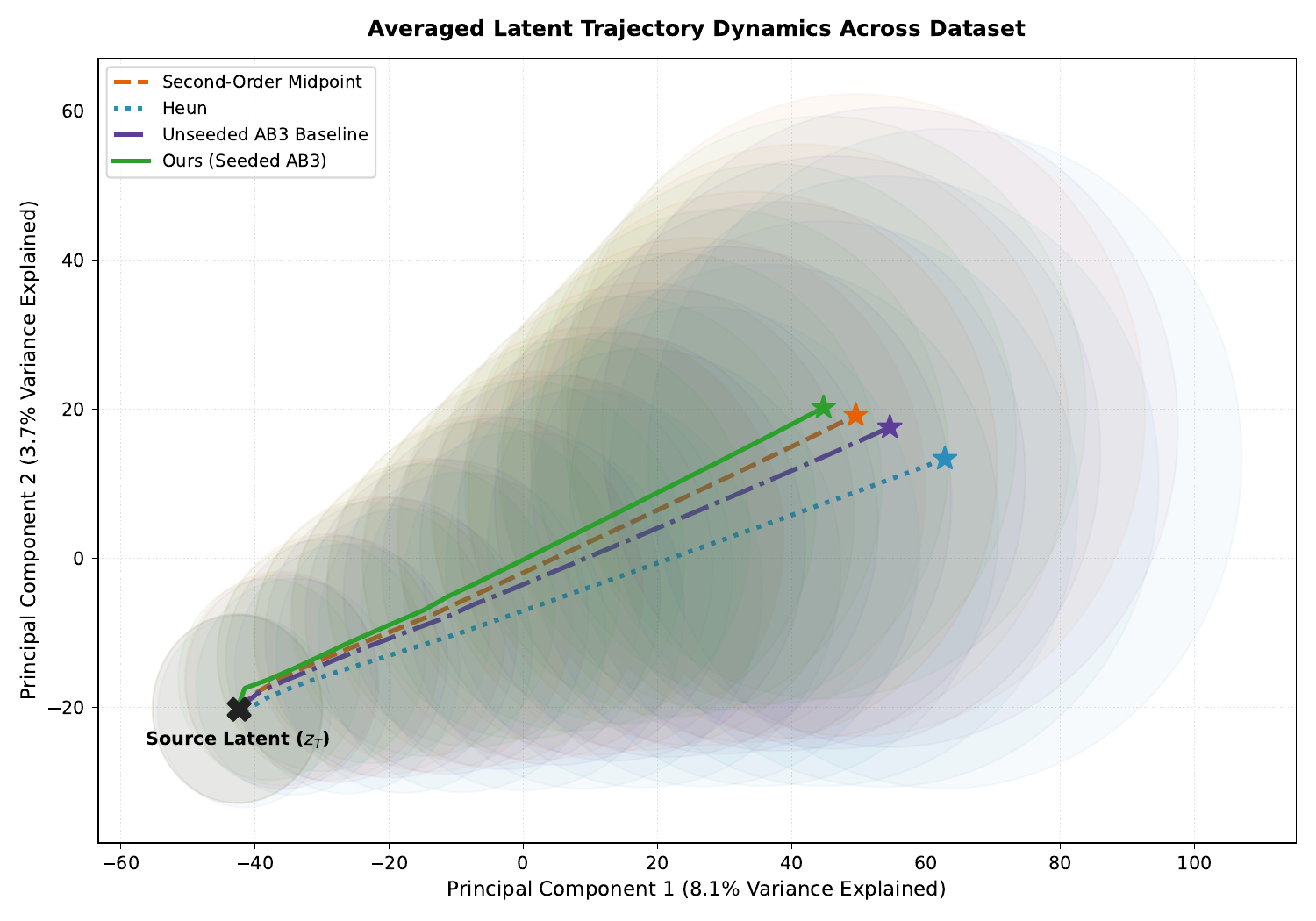}
\caption{Global latent-space trajectory visualization for the timbre-transfer task projected using Principal Component Analysis (PCA). The axes correspond to the two principal directions of variation across the dataset. The shared initial latent state ($\mathbf{z}_T$) is marked by a black $\mathbf{\times}$, while final integration endpoints are denoted by stars. Standard-deviation contours illustrate the spatial distribution of trajectories across different musical examples. Compared with conventional initialization strategies, the proposed Seeded AB3 method exhibits reduced curvature near the initialization boundary and follows a trajectory that more closely aligns with the expected rectified-flow transport direction. The corresponding trajectory visualization for the \textbf{genre-transfer} task is provided in the Appendix.}
\label{fig:trajectory_map_timbre}
\end{figure*}

Although the deterministic inversion described in the previous section provides an accurate latent representation of the input recording, the subsequent editing process is considerably more sensitive to numerical integration errors. During generation, large classifier-free guidance scales are combined with cross-attention feature injection, causing small discretization errors to accumulate rapidly along the latent trajectory. These errors may appear as reconstruction artifacts, degraded timbral consistency, or loss of musical structure.

The deterministic inversion described in the previous section employs the first-order Euler discretization in Eq.~(\ref{eq:euler-step}), which is also the default sampling strategy adopted by FluxMusic~\cite{fluxplaysmusic}. Although computationally efficient, first-order integration has relatively large local truncation error and becomes increasingly inaccurate under strong classifier-free guidance and feature injection during editing.

To improve numerical accuracy, higher-order single-step solvers are employed. In this work, we evaluate both the explicit midpoint method (Second-Order Euler) and Heun's method. The Second-Order Euler solver first estimates the velocity at the midpoint,

\begin{equation}
\mathbf{z}_{k+\frac12}
=
\mathbf{z}_k
+
\frac{\Delta t}{2}
v_\theta(\mathbf{z}_k,t_k),
\end{equation}

and then performs the update

\begin{equation}
\mathbf{z}_{k+1}
=
\mathbf{z}_k
+
\Delta t\,
v_\theta
\!\left(
\mathbf{z}_{k+\frac12},
t_k+\frac{\Delta t}{2}
\right).
\end{equation}

Heun's method instead evaluates the velocity at both the beginning and the end of the interval,

\begin{equation}
\tilde{\mathbf{z}}_{k+1}
=
\mathbf{z}_k
+
\Delta t\,
v_\theta(\mathbf{z}_k,t_k),
\end{equation}

followed by

\begin{equation}
\mathbf{z}_{k+1}
=
\mathbf{z}_k
+
\frac{\Delta t}{2}
\left[
v_\theta(\mathbf{z}_k,t_k)
+
v_\theta(\tilde{\mathbf{z}}_{k+1},t_{k+1})
\right].
\end{equation}

Both methods achieve second-order accuracy while requiring only information from the current integration step, making them straightforward replacements for Euler without requiring any trajectory history.

To further reduce discretization error, we additionally adopt the third-order Adams--Bashforth (AB3) multi-step solver. Unlike single-step methods, AB3 estimates the current update using velocity evaluations from multiple preceding timesteps,

\begin{equation}
\label{eq:ab3-core}
\begin{split}
\mathbf{z}_{k+1}
=
\mathbf{z}_k
+
\Delta t
\Big[
&
\frac{23}{12}
v_\theta(\mathbf{z}_k,t_k)
-
\frac{16}{12}
v_\theta(\mathbf{z}_{k-1},t_{k-1})
\\
&
+
\frac{5}{12}
v_\theta(\mathbf{z}_{k-2},t_{k-2})
\Big].
\end{split}
\end{equation}

Compared with Euler and second-order Runge--Kutta methods, AB3 provides a higher-order approximation of the underlying continuous trajectory by exploiting information accumulated over multiple previous timesteps, reducing the local truncation error from $\mathcal{O}(h^2)$ for Euler and $\mathcal{O}(h^3)$ for second-order methods to $\mathcal{O}(h^4)$.

Although AB3 provides a more accurate approximation of the underlying rectified-flow dynamics than first- and second-order methods, it cannot be applied at the beginning of a trajectory because it requires velocity evaluations from previous integration steps. At the start of generation, this history is unavailable,

\begin{equation}
\mathcal{H}_0
=
\left\{
v_\theta(\mathbf{z}_{-1},t_{-1}),
v_\theta(\mathbf{z}_{-2},t_{-2})
\right\}
=
\emptyset.
\end{equation}

Consequently, conventional implementations initialize the solver using one or more lower-order integration steps, such as Euler or Heun, before switching to AB3. While this warm-start strategy is appropriate for unconditional sampling, it is less suitable for inversion-based editing. The latent representation and cached attention features used during generation originate from the deterministic inversion trajectory, whereas the first generation updates are computed using a different numerical integration scheme. We refer to this mismatch in the numerical treatment of the startup phase between inversion and generation as the \emph{Multi-Step Startup Asymmetry}. Since the Multi-Step Startup Asymmetry occurs precisely at the transition between inversion and generation, the resulting numerical error can propagate throughout the subsequent editing trajectory.

A key observation is that deterministic inversion already evaluates the required velocity field at every integration step. Instead of reconstructing the missing history through lower-order warm-start iterations, we reuse the velocity evaluations obtained during inversion to initialize the multi-step solver.

To overcome this startup limitation of the proposed AB3 integration framework, we further introduce \emph{Dynamic History Caching}. During inversion, the predicted velocity at each integration step is recorded,

\begin{equation}
\mathbf{v}_k
=
v_\theta(\mathbf{z}_k,t_k),
\end{equation}

while only the two most recent velocity evaluations are retained to initialize the generation process. At the first generation step, the current velocity is evaluated in the standard manner, whereas the cached inversion velocities provide the historical information required by AB3. The first update therefore becomes

\begin{equation}
\mathbf{z}_{1}
=
\mathbf{z}_{0}
+
\Delta t
\left(
\frac{23}{12}\mathbf{v}_{0}
-
\frac{16}{12}\mathbf{v}_{\mathrm{inv}}^{K}
+
\frac{5}{12}\mathbf{v}_{\mathrm{inv}}^{K-1}
\right),
\end{equation}

where $\mathbf{v}_0$ is the velocity evaluated at the current generation step, and $\mathbf{v}_{\mathrm{inv}}^{K}$ and $\mathbf{v}_{\mathrm{inv}}^{K-1}$ denote the final two velocity evaluations obtained during deterministic inversion. Consequently, AB3 can be applied immediately without performing lower-order warm-start iterations during generation.

From an algorithmic perspective, DHC modifies only the initialization of the multi-step solver. Since the cached velocities are already computed during deterministic inversion, the proposed strategy introduces no additional neural-network evaluations and incurs only negligible memory overhead while preserving the computational efficiency of the original editing pipeline.

To fairly compare different numerical integration strategies, we further introduce the \emph{Equitonal Transfer} protocol. Different solvers can produce different semantic editing strengths under the same classifier-free guidance (CFG) scale, making direct comparison potentially misleading. We therefore adjust the CFG scale slightly for each solver so that all methods achieve a comparable level of semantic transformation, measured using CLAP similarity. All remaining hyperparameters, including the number of integration steps, feature-injection settings, and model configuration, are kept identical. This normalization ensures that differences in reconstruction quality and preservation of the original recording can be attributed primarily to the numerical integration strategy rather than unequal editing strength.

Figure~\ref{fig:trajectory_map_timbre} provides qualitative evidence for the proposed initialization strategy by visualizing latent trajectories during the timbre-transfer task. The Second-Order Euler and Heun solvers exhibit noticeable curvature immediately after departing from the initial latent state, while the conventional AB3 solver displays additional deviation caused by its lower-order warm-start phase. In contrast, DHC enables AB3 to begin with a consistent derivative history, producing a smoother transition from inversion to generation and a trajectory that remains more closely aligned with the approximately straight transport paths learned by rectified-flow models. A corresponding trajectory analysis for the genre-transfer task is provided in the Appendix, demonstrating that the same behavior generalizes to larger semantic transformations.

The qualitative observations from the trajectory analysis are further supported by the experimental results presented in Section~\ref{sec:results}. In addition to trajectory visualizations, we compare mel-spectrogram reconstructions and evaluate different numerical solvers using both objective and subjective metrics. Together, these analyses demonstrate that DHC consistently improves numerical stability, leading to better semantic consistency with the target prompt, stronger preservation of the original recording, and higher overall editing quality without increasing computational complexity.

\subsection{FLUX that Plays Music}

Recent advances have demonstrated that rectified-flow models are not limited to image synthesis but can also be effectively applied to music generation. \textit{FLUX that Plays Music}~\cite{fluxplaysmusic} extends the FLUX rectified-flow transformer to the text-to-music setting by performing generation in a compressed latent mel-spectrogram space, illustrating the flexibility of rectified-flow architectures across different data modalities.

The framework first transforms an input waveform into a mel-spectrogram, which is subsequently encoded into a compact latent representation using a variational autoencoder (VAE). The rectified-flow transformer operates entirely within this latent space. Text conditioning is obtained from pretrained language and audio-text encoders, including T5~\cite{chung2024scaling}, which provides rich semantic text embeddings, and CLAP~\cite{clap}, whose embeddings are jointly aligned with textual and audio representations. The transformer backbone consists of alternating \emph{double-stream} and \emph{single-stream} blocks. In the double-stream stage, text embeddings and music latents are processed independently while exchanging information through cross-attention, allowing textual descriptions to influence the evolving musical representation. The subsequent single-stream blocks fuse the two modalities by concatenating their token representations, enabling more direct interaction between textual and audio features. In addition, global conditioning information, including prompt-level and temporal embeddings, is incorporated through modulation layers that adaptively scale the hidden representations.

During inference, sampling begins from a Gaussian latent variable $z(0)$ and integrates the learned rectified-flow dynamics until reaching the generated latent representation $z(1)$. The resulting latent is decoded by the VAE into a mel-spectrogram, which is finally converted into an audible waveform using a neural vocoder. Owing to the nearly linear transport trajectories learned by rectified flow, FLUX that Plays Music requires substantially fewer numerical integration steps than conventional diffusion-based text-to-audio models, resulting in more efficient inference while maintaining high synthesis quality.

Despite these advantages, adapting this architecture to inverse problems reveals a critical vulnerability. While the straightened transport trajectories of rectified flow support deterministic backward integration, the highly dense, multi-instrumental nature of audio latents causes standard low-order single-step solvers to accumulate massive discretization errors during inversion. When these errors are compounded by aggressive classifier-free guidance during generation, the trajectory drifts violently away from the continuous probability path, manifesting acoustically as robotic distortions, phase cancellations, and a total loss of melodic structure. Overcoming this path drift without inducing catastrophic VRAM overhead or disrupting the model's internal feature alignment represents a fundamental barrier to high-fidelity audio editing.

\subsection{Encoder}
\label{subsec:encoder}

Our framework employs the pretrained variational autoencoder (VAE) introduced in AudioLDM2~\cite{liu2023audioldm} to encode audio signals into a compact latent representation. Given an input waveform, a TacotronSTFT-based frontend first transforms it into a mel-spectrogram that captures its spectral and temporal characteristics. The mel-spectrogram is subsequently compressed by the VAE into a lower-dimensional latent space, preserving the semantic content of the audio while substantially reducing its dimensionality. All subsequent generation and editing operations are performed in this continuous latent space, which provides an efficient and semantically meaningful representation of the input audio.

\subsection{Attention Feature Reuse}

To preserve the structural characteristics of the source recording during editing, we reuse intermediate attention representations extracted during deterministic inversion. Following prior feature injection and attention-control approaches~\cite{crossattentioncontrol, plugandplay}, we investigate how replacing different attention components affects the trade-off between structural preservation and semantic controllability.

Our modifications are applied only to the single-stream transformer blocks, where text and audio tokens interact through a unified self-attention mechanism. During inversion, we cache the key and value tensors,
$\{\widetilde{\mathcal{K}}_{t_k}^{m}\}$ and
$\{\widetilde{\mathcal{V}}_{t_k}^{m}\}$,
from the last $M$ single-stream blocks over the final $n$ inversion timesteps:

\begin{equation}
\widetilde{\mathbf{F}}_{t_k}^{m}
=
\mathrm{Attention}
(
\widetilde{\mathcal{Q}}_{t_k}^{m},
\widetilde{\mathcal{K}}_{t_k}^{m},
\widetilde{\mathcal{V}}_{t_k}^{m}
),
\end{equation}

where $m$ indexes the selected transformer blocks and $t_k$ denotes the cached inversion timesteps.

During generation, the standard attention operation

\begin{equation}
\mathbf{F}_{t_k}^{m}
=
\mathrm{Attention}
(
\mathcal{Q}_{t_k}^{m},
\mathcal{K}_{t_k}^{m},
\mathcal{V}_{t_k}^{m}
),
\end{equation}

is modified using one of three feature reuse strategies.

\begin{itemize}

\item \textbf{Value Replacement.}
Only the value tensor is replaced:

\begin{equation}
\mathbf{F}_{t_k}^{m'}
=
\mathrm{Attention}
(
\mathcal{Q}_{t_k}^{m},
\mathcal{K}_{t_k}^{m},
\widetilde{\mathcal{V}}_{t_k}^{m}
),
\end{equation}

allowing the attention weights to remain adaptive while preserving the source feature representations~\cite{crossattentioncontrol,plugandplay}.

\item \textbf{Key Replacement.}
Only the key tensor is replaced:

\begin{equation}
\mathbf{F}_{t_k}^{m'}
=
\mathrm{Attention}
(
\mathcal{Q}_{t_k}^{m},
\widetilde{\mathcal{K}}_{t_k}^{m},
\mathcal{V}_{t_k}^{m}
),
\end{equation}

which constrains the attention routing to follow the structural organization captured during inversion~\cite{manifoldrepresentationkeyvision}.

\item \textbf{Key--Value Replacement.}
Both key and value tensors are replaced:

\begin{equation}
\mathbf{F}_{t_k}^{m'}
=
\mathrm{Attention}
(
\mathcal{Q}_{t_k}^{m},
\widetilde{\mathcal{K}}_{t_k}^{m},
\widetilde{\mathcal{V}}_{t_k}^{m}
),
\end{equation}

thereby preserving both the attention structure and the underlying feature representations of the source recording.

\end{itemize}

Section~\ref{sec:results} compares these three strategies and shows that Key--Value replacement provides the best balance between structural preservation and text-guided editability. We therefore adopt this configuration in all subsequent experiments.

\begin{figure*}[!t]
  \centering
  \includegraphics[width=\textwidth]{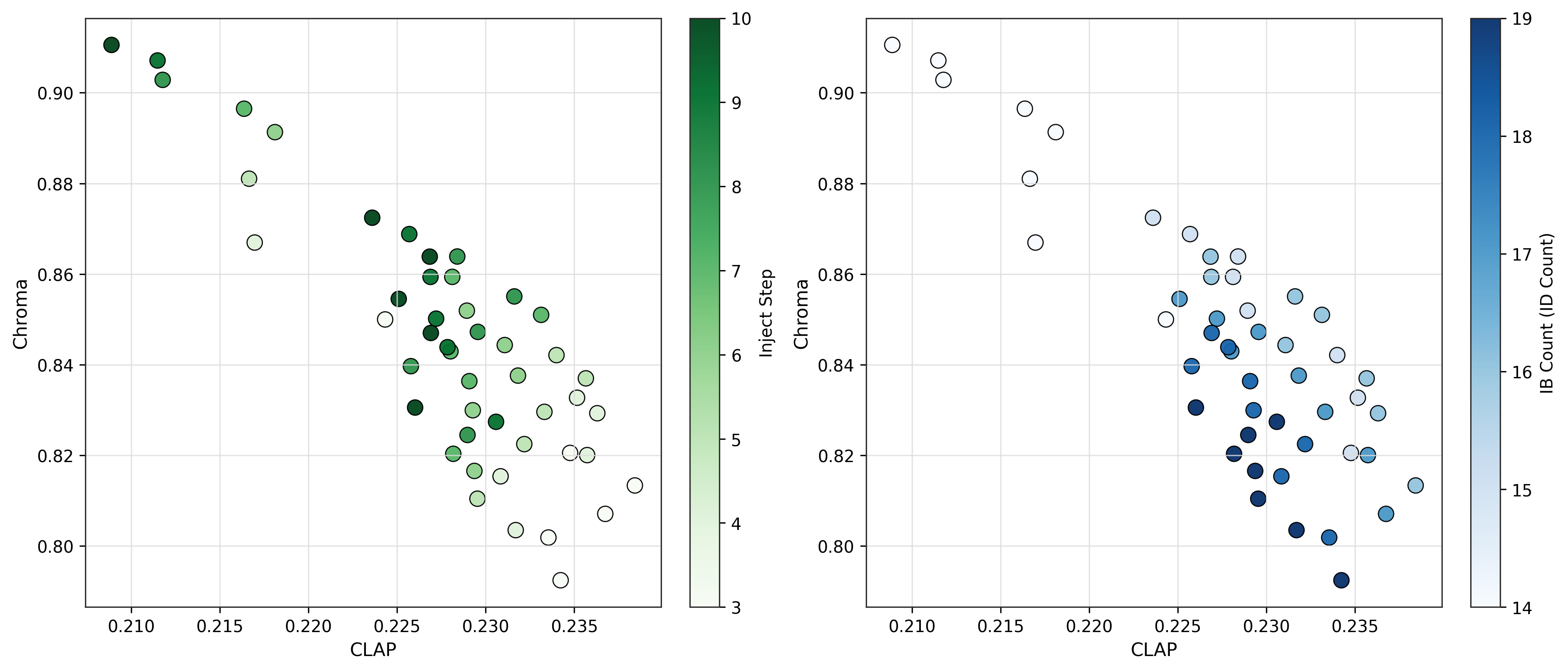}
   \caption{Influence of injection steps and injection block (IB) count on the transferability--fidelity trade-off for the timbre transfer task. The figure illustrates the effect of injecting the attention value (V) representations during generation on the balance between semantic transferability and fidelity to the source audio. Results for key (K) injection, combined key-and-value (K+V) injection, and the corresponding experiments on genre transfer are provided in the Appendix.}
  \label{fig:comparison}
\end{figure*}

\vspace{-0.1cm}
\setlength{\abovedisplayskip}{6pt}
\setlength{\belowdisplayskip}{6pt}

\section{Experiments}
\label{sec:experiments}

\subsection{Datasets}

We construct two curated evaluation datasets, each containing 40 music clips collected from publicly available YouTube recordings. One dataset is designed for timbre transfer, while the other focuses on genre transformation. Each clip was manually selected to ensure a clear distinction between instrument or genre categories while minimizing background noise and recording artifacts. All audio samples were resampled to 16~kHz and either trimmed or segmented into clips of uniform 10-second duration.

Although the evaluation datasets are relatively small, they are designed for controlled benchmarking rather than large-scale model training. Since every editing method is evaluated on exactly the same set of input recordings and text prompts, the datasets provide a fair and consistent basis for comparing different editing algorithms. Moreover, the selected clips cover a diverse range of musical characteristics, enabling evaluation across different instruments and musical styles.

The timbre-transfer dataset contains recordings of five instruments: \textit{electric guitar}, \textit{flute}, \textit{piano}, \textit{violin}, and \textit{acoustic guitar}. These instruments exhibit diverse harmonic, percussive, and timbral characteristics, providing a representative benchmark for evaluating instrument transformation while preserving the musical content of the source recording.

The genre-transfer dataset consists of music from four representative styles: \textit{pop}, \textit{jazz}, \textit{rock}, and \textit{hip-hop}. These genres span substantially different rhythmic, harmonic, and stylistic characteristics, allowing evaluation of the model's ability to perform high-level semantic edits while preserving the identity of the original recording.

\subsection{Baselines}
\label{subsec:baselines}

We compare the proposed method against four representative text-to-music generation and editing approaches.

\begin{itemize}

\item \textbf{AudioLDM2:}
\textit{AudioLDM2}~\cite{audioldm2} is a latent diffusion model for text-conditioned audio generation that employs CLAP and Flan-T5 embeddings together with a U-Net architecture incorporating cross-attention. For editing, we adopt an SDEdit-style procedure~\cite{meng2021sdedit}, where the source audio is partially diffused to an intermediate timestep $t_{\text{edit}} < T$. Reverse diffusion is then performed under the guidance of the target text prompt to obtain the edited audio.

\item \textbf{MusicGen:}
We additionally evaluate \textit{MusicGen}~\cite{musicgenstem}, an autoregressive Transformer that generates discrete audio tokens rather than latent diffusion representations. Specifically, we use the \textit{MusicGen-Melody (1.5B)} variant, which supports melody conditioning through chromagram features. During evaluation, the source chromagram serves as the melody condition while the editing instruction is provided as the text prompt, producing the edited output $\tilde{x}$.

\item \textbf{ZETA (DDPM Inversion):}
\textit{Zero-Shot Unsupervised and Text-Based Audio Editing Using DDPM Inversion}~\cite{zeta} performs editing by first inverting the source audio into the latent space of a diffusion model and subsequently steering the reverse denoising process using textual guidance. We employ the ZETA variant as a representative diffusion-based inversion baseline for zero-shot music editing.

\item \textbf{FluxMusic:}
\textit{FluxMusic}~\cite{fluxplaysmusic} extends large-scale rectified-flow transformers to text-to-music generation. To adapt it for editing, we first invert the source recording into the latent space using deterministic first-order Euler integration and then perform a forward generation pass conditioned on the target prompt. This baseline provides a direct comparison with a standard rectified-flow editing pipeline that does not employ the proposed history-seeded multi-step integration strategy.

\end{itemize}

We also examined several additional methods, including \textit{MusicMagus}~\cite{musicmagus}, \textit{TransPlayer}~\cite{transplayer}, \textit{MelodyFlow}~\cite{melodyflow}, and \textit{SteerMusic}~\cite{steermusic}. However, these approaches were excluded from the quantitative comparison for practical reasons. \textit{MusicMagus} exhibited limited performance on real-world music editing, while \textit{TransPlayer} supports only a limited set of editing operations that do not align with our evaluation protocol. Finally, at the time of this study, neither \textit{MelodyFlow} nor \textit{SteerMusic} had publicly released code or pretrained checkpoints, preventing a fair and reproducible evaluation.

\begin{figure*}[!t]
  \centering
  \begin{minipage}[b]{0.24\textwidth}
    \centering
    \includegraphics[width=\textwidth]{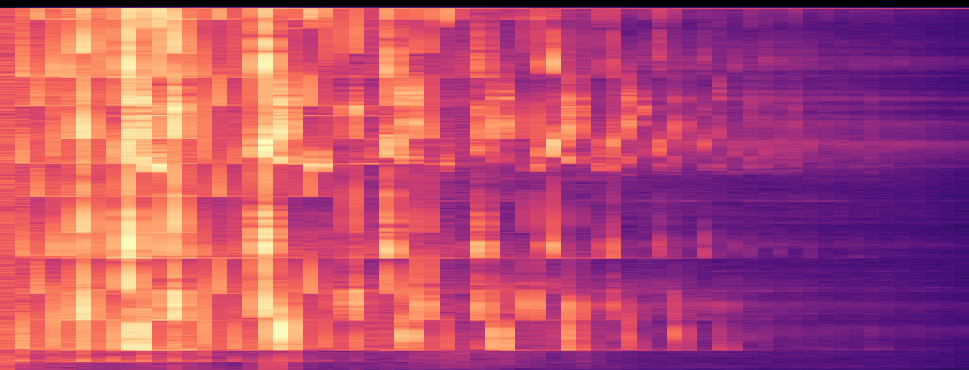}
  \end{minipage}
  \hfill
  \begin{minipage}[b]{0.24\textwidth}
    \centering
    \includegraphics[width=\textwidth]{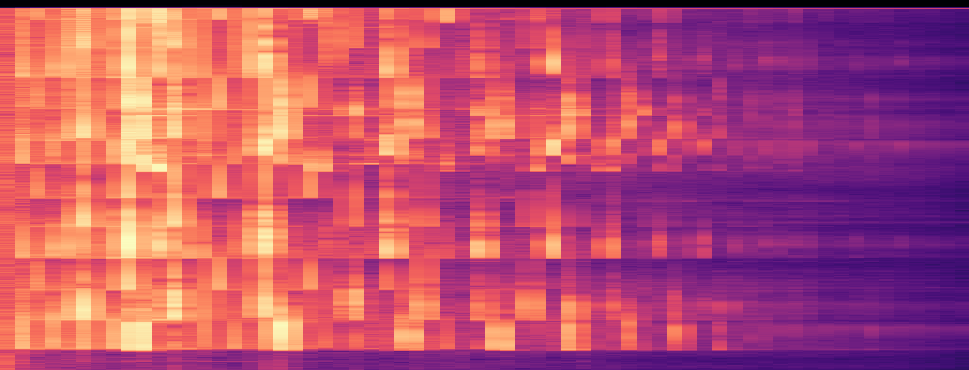}
  \end{minipage}
  \hfill
  \begin{minipage}[b]{0.24\textwidth}
    \centering
    \includegraphics[width=\textwidth]{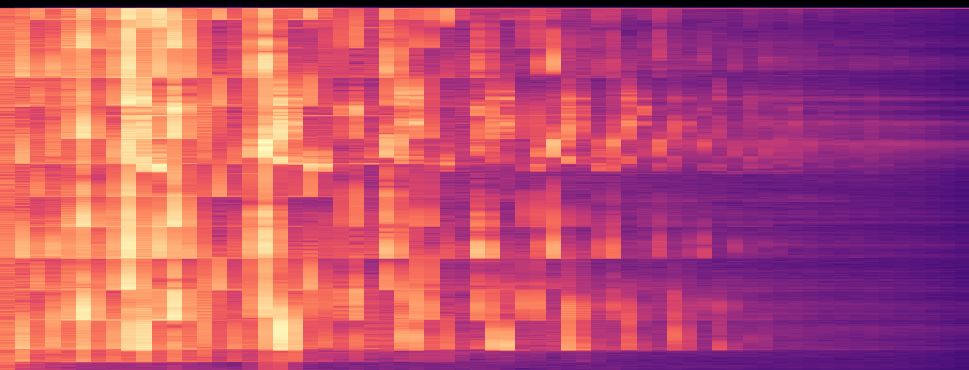}
  \end{minipage}
  \hfill
  \begin{minipage}[b]{0.24\textwidth}
    \centering
    \includegraphics[width=\textwidth]{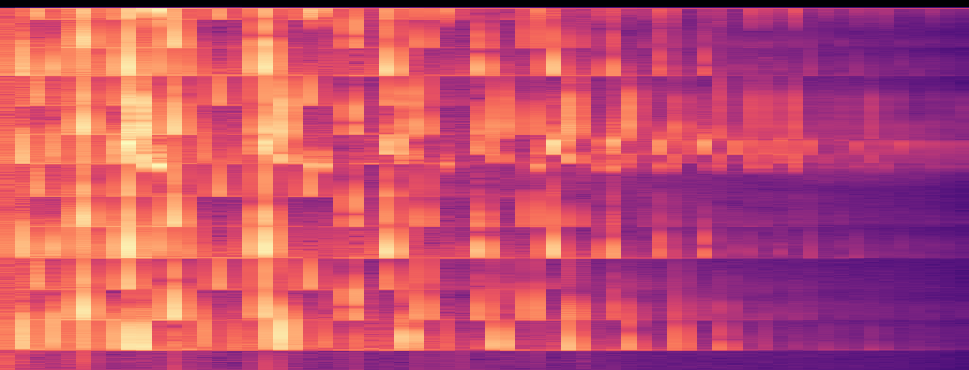}
  \end{minipage}
  
  \caption{Qualitative comparison of Mel spectrograms generated using different numerical integration strategies within the proposed editing framework. From left to right: Heun, Second-Order Euler, conventional unseeded third-order Adams--Bashforth (AB3), and the proposed Seeded AB3 solver. Although all methods preserve the overall time--frequency structure of the generated audio, differences can be observed in the distribution and continuity of spectral energy. Compared with the conventional AB3 solver, the proposed Seeded AB3 method produces a more uniform spectral energy distribution with fewer localized variations, suggesting improved numerical stability through the reuse of inversion history while maintaining the same editing behavior.}
\label{fig:solver_spectrogram_ablation}
\end{figure*}

\begin{figure*}[!t]
  \centering
  \begin{minipage}[b]{0.32\textwidth}
    \centering
    \includegraphics[width=\textwidth]{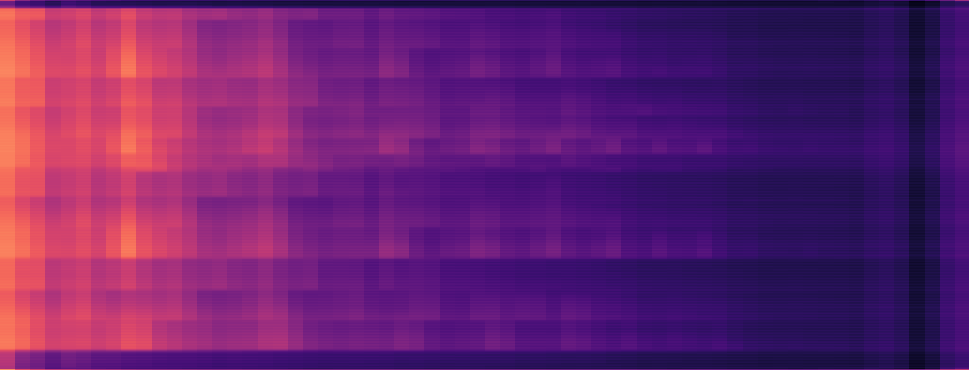}
  \end{minipage}
  \hfill
  \begin{minipage}[b]{0.32\textwidth}
    \centering
    \includegraphics[width=\textwidth]{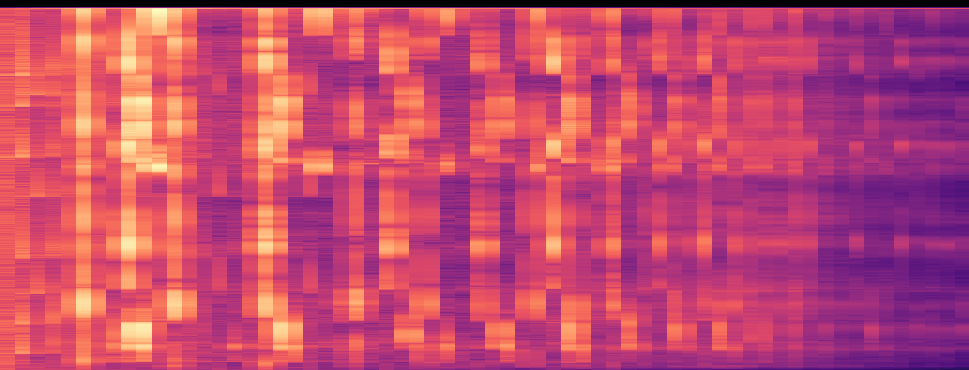}
  \end{minipage}
  \hfill
  \begin{minipage}[b]{0.32\textwidth}
    \centering
    \includegraphics[width=\textwidth]{figures/piano_3_violin_ours.png}
  \end{minipage}
  
  \caption{Qualitative comparison of Mel spectrograms generated using different numerical integration strategies within the FluxMusic editing framework. \textbf{Left:} The standard Euler solver produces limited harmonic development and weaker high-frequency content, reflecting conservative editing behavior. \textbf{Center:} Replacing Euler with an unseeded third-order Adams--Bashforth (AB3) solver increases editing strength but introduces noticeable spectral artifacts, including smeared harmonic structures and diffuse background energy caused by unstable multi-step initialization. \textbf{Right:} The proposed Seeded AB3 framework with DHC suppresses these initialization artifacts, producing cleaner harmonic ridges, sharper time--frequency structures, and a more coherent energy distribution throughout the generated audio.}

\label{fig:fluxmusic_spectrogram_ablation}
\end{figure*}

\subsection{Objective Metrics}
\label{subsec:objective_metrics}

We assess the proposed method using five complementary objective metrics that jointly evaluate semantic consistency, structural preservation, spectral similarity, and perceptual quality.

\begin{itemize}

\item \textbf{CLAP Similarity:}
CLAP~\cite{clap} projects both audio and text into a shared embedding space learned through contrastive training. We compute the cosine similarity between the embedding of the edited audio $\tilde{x}$ and that of the target prompt $y$, where larger values indicate stronger semantic correspondence.

\item \textbf{Chroma Similarity:}
To measure structural preservation, we compute the chroma similarity between the source audio $x$ and the edited audio $\tilde{x}$. Chroma features are extracted using the Constant-Q Transform (CQT) implementation provided by \texttt{librosa}~\cite{mcfee2015librosa}. Frame-wise cosine similarity between the resulting chromagrams quantifies how effectively harmonic and rhythmic structures are retained after editing.

\item \textbf{CQT-PCC:}
The Constant-Q Transform (CQT)~\cite{brown1991cqt} represents audio on a logarithmic frequency scale that closely matches human pitch perception. We compute the Pearson Correlation Coefficient (PCC) between the CQT magnitude spectra of the original audio $x$ and the edited audio $\tilde{x}$. Higher correlation values indicate better preservation of harmonic and timbral content.

\item \textbf{Fréchet Audio Distance (FAD):}
Perceptual quality is evaluated using the Fréchet Audio Distance (FAD), which measures the distributional distance between real and generated audio embeddings extracted by a pretrained VGGish network. Similar to the Fréchet Inception Distance (FID) commonly used in image generation, FAD compares multivariate Gaussian distributions fitted to these embeddings. Lower FAD values indicate that the generated audio is perceptually closer to real recordings.

\end{itemize}

\subsection{Subjective Metrics}
\label{subsec:subjective_metrics}

In addition to objective evaluation, we conduct a listening study to assess the perceptual quality of the edited music. Following the ITU-T recommendations for multimedia quality assessment~\cite{itut1999video,itut1996audio}, participants evaluate each sample using two Mean Opinion Score (MOS) criteria: \textit{MOS-T} and \textit{MOS-P}.

\textbf{MOS-T} measures the semantic consistency between the generated music and the target text prompt. Participants assign scores on a five-point Likert scale according to how accurately the edited audio reflects the intended musical style, emotion, and semantic content described by the prompt.

\textbf{MOS-P} evaluates the perceptual similarity between the edited audio $\tilde{x}$ and the original recording $x$. Participants assess how well the edited result preserves the timbre, rhythmic structure, and overall musical characteristics of the source while naturally incorporating the requested modifications. Higher MOS-P scores indicate stronger preservation of the original recording.

\subsection{Hyperparameter Selection}

The quality of the edited music is influenced by several inference-time hyperparameters. To better understand their effects, we performed a series of experiments investigating different parameter configurations. In total, our framework exposes five tunable hyperparameters: the number of diffusion timesteps, the target classifier-free guidance (CFG) scale, the source CFG scale, the number of injection steps, and the injection block count (IB count).

An exhaustive search over all five parameters would be computationally prohibitive. Therefore, we perform a detailed analysis of the injection steps and IB count, while selecting the remaining hyperparameters empirically. Across all editing tasks, we found that target CFG values between 5 and 15 consistently provide strong semantic editing performance, whereas fixing the source CFG to 1 yields stable inversion. We further set the number of diffusion timesteps to 25. Although additional timesteps generally improve generation quality, 25 steps provide a favorable compromise between editing performance and computational efficiency. In comparison, the original FluxMusic model employs 50 diffusion timesteps during generation.

The \textbf{injection step} determines the diffusion timesteps at which cross-attention features cached during inversion are reintroduced into the generation process (Figure~\ref{fig:comparison}). Increasing the number of injection steps generally improves reconstruction fidelity, but at the expense of editing strength. When too many injection steps are used, the generated audio preserves excessive acoustic details from the source recording, reducing its ability to follow the target editing instruction.

The \textbf{IB count} specifies the single transformer block after which attention injection is performed within each selected diffusion step. Specifically, if the transformer contains \(n\) single-stream blocks and the IB count is set to \(m\) (\(m<n\)), feature injection is applied immediately after the \(m\)-th block. As illustrated in Figure~\ref{fig:comparison}, larger IB counts typically improve semantic transferability while reducing reconstruction fidelity, highlighting the inherent trade-off between preserving the source recording and applying the desired edit.

\section{Results and Discussion}
\label{sec:results}

\subsection{Time--Frequency Analysis of the Proposed Framework}
\label{subsec:solver_metrics_analysis}

While the quantitative metrics presented later summarize the overall editing performance, they provide limited insight into how individual components of the proposed framework affect the spectral structure of the generated audio. To complement the quantitative evaluation, we qualitatively compare Mel spectrograms generated by different variants of the editing pipeline. Mel spectrograms provide an intuitive visualization of harmonic organization, spectral energy distribution, and time--frequency artifacts introduced during the editing process.

Figure~\ref{fig:solver_spectrogram_ablation} compares the Mel spectrograms generated using four numerical integration strategies. Although all methods preserve the overall harmonic structure of the generated audio, noticeable differences emerge in the distribution and continuity of spectral energy. Compared with the second-order solvers, both AB3 variants produce richer and more coherent harmonic patterns. More importantly, the proposed Seeded AB3 solver exhibits a more uniform and continuous energy distribution than the conventional unseeded AB3 solver, with fewer localized irregularities across the time--frequency representation. These qualitative observations are consistent with the proposed DHC mechanism, suggesting that reusing the inversion history improves the numerical stability of high-order integration while preserving the desired musical structure.

\begin{table*}[t]
\centering
\caption{Quantitative evaluation of numerical ODE solver trajectories on the timbre-transformation task.}
\label{tab:solver_metrics}
\begin{tabular}{lcccc}
\hline
\textbf{Solver Configuration} & \textbf{CLAP} $\uparrow$ & \textbf{Chroma} $\uparrow$ & \textbf{FAD} $\downarrow$ & \textbf{CQT-PCC} $\uparrow$ \\
\hline
Heun & .233 & .812 & 4.121 & .558 \\
Second-Order Euler & .237 & .836 & 4.456 & .596 \\
Unseeded AB3 & .236 & .842 & \textbf{3.869} & .612 \\
\textbf{Ours (Seeded AB3)} & \textbf{.238} & \textbf{.860} & 3.938 & \textbf{.638} \\
\hline
\end{tabular}
\label{tab:solver_metrics_timbre}
\end{table*}

\begin{table*}[!t]
\centering
\caption{Quantitative evaluation of numerical ODE solver trajectories on the genre-transformation task.}
\begin{tabular}{lcccc}
\hline
\textbf{Solver Configuration} & \textbf{CLAP} $\uparrow$ & \textbf{Chroma} $\uparrow$ & \textbf{FAD} $\downarrow$ & \textbf{CQT-PCC} $\uparrow$ \\
\hline
Heun               & .534          & .771          & 6.590 & .359 \\
Second-Order Euler & .536          & .784          & 6.172          & .408 \\
Unseeded AB3       & .531          & .786          & 5.674          & .420  \\
\textbf{Ours (Seeded AB3)} & \textbf{.538} & \textbf{.797} & \textbf{4.693} & \textbf{.468} \\  
\hline
\end{tabular}
\label{tab:solver_metrics_genre}
\end{table*}

Figure~\ref{fig:fluxmusic_spectrogram_ablation} further examines the contribution of cross-attention feature injection. Compared with the Seeded-AB3 solver alone, incorporating cross-attention feature injection produces only minor changes in the overall spectral appearance. This observation is expected, as the primary purpose of feature injection is to preserve the structural characteristics of the source recording rather than to alter the spectral distribution of the generated audio. Its benefits are therefore reflected more clearly in the structural preservation metrics presented in the following subsection.

Taken together, these qualitative observations reinforce the complementary roles of the two proposed components. DHC primarily improves numerical stability during high-order integration, whereas cross-attention feature injection preserves structural consistency throughout the editing process. These findings are consistent with both the latent trajectory analysis and the quantitative evaluation reported in the following subsection.

\subsection{Numerical Solver Ablation}
\label{subsec:solver_ablation}

To evaluate the effect of high-order multi-step integration together with the proposed DHC mechanism, we conduct an ablation study comparing several numerical ODE solvers within the same editing framework. To ensure a controlled comparison, all configurations employ the same Cross-Attention Key--Value (KV) feature-injection backbone. We consider four solver variants: Heun, Second-Order Euler, Unseeded Adams--Bashforth 3 (AB3), and the proposed Seeded AB3 method. The evaluation is performed on both timbre-transfer and genre-transformation tasks to investigate whether the proposed initialization strategy consistently improves editing quality across different levels of musical transformation.

The quantitative results are summarized in Tables~\ref{tab:solver_metrics_timbre} and~\ref{tab:solver_metrics_genre}. Using the Equitonal Transfer protocol, all solver configurations operate at comparable semantic editing depths by slightly adjusting the classifier-free guidance scale for each solver while keeping all remaining hyperparameters and experimental settings identical. Consequently, differences in harmonic preservation, structural consistency, and perceptual quality can be attributed primarily to the numerical integration strategy rather than unequal editing strength.

For the timbre-transfer task, the Heun solver produces the weakest overall performance, obtaining the lowest Chroma similarity (0.812) and CQT-PCC (0.558), despite achieving competitive semantic alignment. Second-Order Euler improves both harmonic preservation and structural consistency, increasing Chroma similarity to 0.836 and CQT-PCC to 0.596, although this comes at the expense of the highest Fréchet Audio Distance (FAD = 4.456). The unseeded AB3 solver further benefits from higher-order integration, achieving the lowest FAD (3.869) while improving both Chroma similarity and CQT-PCC over the second-order methods. Building upon this stronger baseline, the proposed Seeded AB3 method achieves the highest CLAP similarity (0.238), Chroma similarity (0.860), and CQT-PCC (0.638), while maintaining a competitive FAD (3.938). These results indicate that reusing the cached inversion history consistently improves semantic alignment and structural preservation without sacrificing perceptual quality.

A similar trend is observed for the more challenging genre-transformation task, demonstrating that the proposed initialization strategy generalizes beyond local timbral modifications. Heun again exhibits the weakest structural preservation, yielding the lowest Chroma similarity (0.771), lowest CQT-PCC (0.359), and highest FAD (6.590). Second-Order Euler improves both harmonic and structural consistency, while the unseeded AB3 solver provides an additional improvement by further reducing perceptual distortion and increasing structural preservation. The proposed Seeded AB3 method consistently achieves the best performance across all evaluated metrics, obtaining the highest semantic alignment (CLAP = 0.538), harmonic similarity (Chroma = 0.797), and structural correlation (CQT-PCC = 0.468), while simultaneously achieving the lowest Fréchet Audio Distance (FAD = 4.693). The consistent improvements across every objective metric demonstrate that DHC effectively stabilizes high-order integration, enabling more accurate semantic editing while better preserving the harmonic and structural characteristics of the original recording.

Overall, the ablation study demonstrates that higher-order integration alone provides clear advantages over second-order solvers, as evidenced by the strong performance of the conventional AB3 method. However, initializing AB3 with the proposed DHC strategy consistently yields further improvements in semantic alignment and structural preservation. The resulting gains are particularly pronounced for the more demanding genre-transfer task, where the proposed Seeded AB3 solver achieves the best performance across all evaluated metrics, confirming that eliminating the warm-start inconsistency leads to a more stable and effective editing process.

\begin{table*}[t]
\centering
\caption{The objective evaluation results on the timbre transfer.}
\begin{tabular}{lcccccc}
\hline
\textbf{Model} & \textbf{Type} & \textbf{CLAP $\uparrow$} & \textbf{Chroma $\uparrow$} & \textbf{CLAP+Chroma Avg. $\uparrow$} & \textbf{CQT-1 PCC $\uparrow$} & \textbf{FAD $\downarrow$} \\
\hline
MusicGen & Supervised  & .220 & .756 & .488 & .278 & 5.343 \\
AudioLDM2 & Zero-shot  & .229 & .817 & .523  & .557 & \textbf{3.623} \\
Zeta & Zero-shot  & .225 & .814 & .519 & .561 & 5.614 \\
FluxMusic & Zero-shot  & -.094 & .600  & .253 & .304 & 18.371 \\
\hline
FlowSonic No Injection\textbf{(ours)} & Zero-shot  & .233 & .770  & .502 & .505 & 5.735 \\
FlowSonic K Injection\textbf{(ours)} & Zero-shot  & \underline{.236} & .783 & .509 & .523 & 5.710 \\
FlowSonic KV Injection\textbf{(ours)} & Zero-shot  & \textbf{.238} & \textbf{.860} & \textbf{.549} & \textbf{.638} & 3.938 \\
FlowSonic V Injection\textbf{(ours)} & Zero-shot  & .231 & \underline{.858} & \underline{.545} & \underline{.637} & \underline{3.887} \\
\hline
\end{tabular}
\label{tab:objective_timbre}
\end{table*}

\begin{table*}[t]
\centering
\caption{The objective evaluation results on the genre transfer.}
\begin{tabular}{lcccccc}
\hline
\textbf{Model} & \textbf{Type} & \textbf{CLAP $\uparrow$} & \textbf{Chroma $\uparrow$} & \textbf{CLAP+Chroma Avg. $\uparrow$} & \textbf{CQT-1 PCC $\uparrow$} & \textbf{FAD $\downarrow$} \\
\hline
MusicGen & Supervised  & .462 & .750 & .606 & .127 & 9.935 \\
AudioLDM2 & Zero-shot  & \textbf{.583} & .694 & .639 & .154 & 8.834 \\
Zeta & Zero-shot & .535 & .763 & .650 & .314 & 6.930 \\
FluxMusic & Zero-shot & .043 & .664 & .354 & .086 & 22.340 \\
\hline
FlowSonic No Injection\textbf{(ours)} & Zero-shot & \underline{.547} & .772 & .660 & .391 & 6.234 \\
FlowSonic K Injection\textbf{(ours)} & Zero-shot & \underline{.547} & .779 & \underline{.663} & .402 & 6.217 \\
FlowSonic KV Injection\textbf{(ours)} & Zero-shot & .538 & \underline{0.797} & \textbf{.667} &  \underline{.468} & \underline{4.693}\\
FlowSonic V Injection\textbf{(ours)} & Zero-shot & .534 & \textbf{.800} & \textbf{.667} & \textbf{.471} & \textbf{4.691} \\
\hline
\end{tabular}
\label{tab:objective_genre}
\end{table*}

\subsection{Objective Results}
\label{subsec:objective_evaluation}

We conduct an objective evaluation to quantitatively assess the performance of the proposed FlowSonic framework on both timbre- and genre-transfer tasks. The evaluation relies on five complementary metrics: CLAP similarity, which measures semantic alignment between the generated and target audio; Chroma similarity, which reflects harmonic fidelity to the source; CLAP+Chroma Avg., which captures the balance between semantic transfer and structural preservation; CQT-1 PCC, which measures spectral correlation; and Fréchet Audio Distance (FAD), which evaluates perceptual realism.

Table~\ref{tab:objective_timbre} presents the quantitative results for the timbre-transfer task. A clear trend emerges when comparing the different stages of our framework. Replacing the original FluxMusic inversion and generation pipeline with the proposed Seeded-AB3 solver already leads to a substantial improvement, as demonstrated by the FlowSonic No Injection variant. This result indicates that a large portion of the performance gain originates from the proposed numerical framework itself. Incorporating cross-attention feature injection further improves the results, confirming that the cached structural representations provide complementary information beyond the benefits of the proposed Seeded-AB3 integration strategy alone.

Among the injection strategies, FlowSonic KV Injection achieves the highest CLAP similarity, indicating the strongest semantic alignment with the target editing prompt. It also obtains the highest Chroma similarity and CQT-1 PCC, demonstrating superior preservation of harmonic content and spectral structure. FlowSonic V Injection consistently ranks second on these metrics, suggesting that preserving the value stream plays a particularly important role in maintaining musical structure during editing. When considering the combined CLAP+Chroma score, KV Injection again provides the best balance between semantic modification and structural fidelity. For perceptual realism, AudioLDM2 achieves the lowest FAD, while FlowSonic V Injection and KV Injection obtain the second- and third-best scores, respectively. Overall, the KV and V Injection variants substantially outperform FluxMusic across all objective metrics, demonstrating that the proposed framework effectively improves both editing accuracy and structural preservation.

Table~\ref{tab:objective_genre} reports the results for the more challenging genre-transfer task. The same progression is consistently observed. Replacing the original FluxMusic inversion and generation procedure with the proposed framework already produces a clear improvement across the objective metrics, demonstrating that the proposed numerical formulation generalizes beyond local timbral modifications. Introducing cross-attention feature injection further improves both structural preservation and semantic consistency, with the KV and V Injection variants consistently outperforming the corresponding no-injection baseline.

For semantic alignment, AudioLDM2 achieves the highest CLAP score, followed closely by FlowSonic No Injection and K Injection, indicating that preserving fewer attention features provides greater flexibility for large-scale stylistic transformations. In contrast, the KV and V Injection variants achieve the highest Chroma similarity and the strongest CQT-1 PCC values, demonstrating superior preservation of harmonic relationships and spectral structure. They also obtain the highest CLAP+Chroma averages, reflecting the best overall balance between semantic editing and musical consistency. In terms of perceptual realism, FlowSonic V Injection achieves the lowest FAD by a substantial margin, indicating that it generates the most realistic audio distribution among all evaluated methods. FlowSonic KV Injection follows closely while maintaining slightly stronger semantic alignment. These results suggest that preserving value representations is particularly effective for maintaining musical coherence during large-scale genre transformations.

Overall, the objective evaluation demonstrates two complementary contributions of the proposed framework. First, the Seeded-AB3 solver substantially improves editing quality compared with the original FluxMusic pipeline, even without feature injection. Second, injecting cached cross-attention representations provides an additional performance gain by strengthening semantic alignment while better preserving harmonic and structural characteristics. Across both timbre- and genre-transfer tasks, the KV and V Injection variants consistently achieve the strongest overall performance, with KV Injection generally providing the best balance between semantic accuracy and structural preservation, while V Injection often achieves the highest perceptual realism.

\begin{table*}[t]
\centering

\begin{minipage}{0.47\textwidth}
\centering
\captionof{table}{Subjective evaluation results on the timbre-transfer task.}
\label{tab:timbre_subjective}
\begin{tabular}{lccc}
\toprule
\textbf{Model} & \textbf{MOS-T $\uparrow$} & \textbf{MOS-P $\uparrow$} & \textbf{Overall $\uparrow$} \\
\midrule
MusicGen                   & \underline{3.75} & 2.20 & 2.98 \\
AudioLDM2                  & 3.25 & 3.20 & 3.23 \\
Zeta                       & 3.25 & \underline{3.45} & 3.35 \\
FluxMusic                  & 1.10 & 1.05 & 1.08 \\
\hline
FlowSonic No Injection (\textbf{ours}) & 3.55 & 3.35 & 3.45 \\
FlowSonic K Injection (\textbf{ours})  & 3.25 & 3.25 & 3.25 \\
FlowSonic KV Injection (\textbf{ours}) & \textbf{4.00} & \textbf{4.20} & \textbf{4.10} \\
FlowSonic V Injection (\textbf{ours})  & \textbf{4.00} & \underline{4.10} & \underline{4.05} \\
\bottomrule
\end{tabular}
\end{minipage}
\hfill
\begin{minipage}{0.47\textwidth}
\centering
\captionof{table}{Subjective evaluation results on the genre-transfer task.}
\label{tab:genre_subjective}
\begin{tabular}{lccc}
\toprule
\textbf{Model} & \textbf{MOS-T $\uparrow$} & \textbf{MOS-P $\uparrow$} & \textbf{Overall $\uparrow$} \\
\midrule
MusicGen                   & 2.45 & 2.05 & 2.25 \\
AudioLDM2                  & \textbf{4.30} & 1.30 & 2.80 \\
Zeta                       & 3.00 & 3.10 & 3.05 \\
FluxMusic                  & 1.00 & 1.00 & 1.00 \\
\hline
FlowSonic No Injection (\textbf{ours}) & 3.50 & 2.95 & 3.23 \\
FlowSonic K Injection (\textbf{ours})  & 3.50 & 3.30 & 3.40 \\
FlowSonic KV Injection (\textbf{ours}) & \underline{3.95} & \underline{4.15} & \underline{4.05} \\
FlowSonic V Injection (\textbf{ours})  & \underline{4.00} & \textbf{4.35} & \textbf{4.18} \\
\bottomrule
\end{tabular}
\end{minipage}

\end{table*}

\subsection{Subjective Results}

To evaluate the perceptual quality of the generated music, we conducted an online listening study using Google Forms with 20 participants, consisting of 10 professional musicians and 10 ordinary listeners without formal musical training. Each participant evaluated one timbre-transfer example and one genre-transfer example. For every sample, participants rated two aspects on a five-point Likert scale: the Mean Opinion Score for Text Alignment (MOS-T), which measures the semantic consistency between the edited music and the target text prompt, and the Mean Opinion Score for Preservation (MOS-P), which evaluates how well the edited audio preserves the timbre, rhythmic structure, and overall musical characteristics of the original recording while naturally incorporating the requested edits. Tables~\ref{tab:timbre_subjective} and~\ref{tab:genre_subjective} summarize the overall results, while a detailed breakdown for professional and ordinary listeners is provided in the Appendix. Overall, the subjective evaluations closely follow the trends observed in the objective metrics, providing additional evidence for the effectiveness of the proposed framework.

Table~\ref{tab:timbre_subjective} presents the subjective evaluation results for the timbre-transfer task. A notable observation is the substantial improvement achieved solely by replacing the original FluxMusic inference procedure with the proposed numerical framework. Even without cross-attention feature injection, \textit{FlowSonic No Injection} increases the overall subjective score from 1.08 to 3.45, demonstrating that the proposed Seeded-AB3 solver together with the DHC strategy substantially improves editing stability and perceptual performance. Introducing cross-attention feature injection provides additional gains. Among all methods, \textit{FlowSonic KV Injection} achieves the highest MOS-T (4.00), MOS-P (4.20), and overall score (4.10), indicating the best balance between semantic consistency with the editing prompt and preservation of the original recording. \textit{FlowSonic V Injection} performs similarly, obtaining the same MOS-T with only a slightly lower MOS-P and overall score. In contrast, \textit{FlowSonic K Injection} provides comparatively smaller improvements, suggesting that injecting only key features is less effective than utilizing value features or their combination. All proposed FlowSonic variants consistently outperform the baseline editing methods.

Table~\ref{tab:genre_subjective} reports the subjective evaluation results for the genre-transfer task. Similar to the timbre-transfer setting, introducing the proposed numerical framework alone leads to a dramatic improvement over the original FluxMusic baseline, increasing the overall subjective score from 1.00 to 3.23 without any feature injection. This demonstrates that a significant portion of the performance gain originates from the proposed numerical solver rather than the attention injection mechanism itself. Incorporating cross-attention feature injection further enhances editing performance. \textit{FlowSonic V Injection} achieves the highest overall score (4.18) together with the highest MOS-P (4.35), indicating that value-feature injection best preserves the musical characteristics of the original recording while successfully incorporating the requested genre modification. \textit{FlowSonic KV Injection} closely follows with an overall score of 4.05 while maintaining strong semantic consistency with the target prompt (MOS-T = 3.95). Although \textit{AudioLDM2} achieves the highest MOS-T (4.30), its very low MOS-P (1.30) indicates that, despite closely following the target prompt, it fails to preserve the characteristics of the original recording, resulting in a considerably lower overall preference than the proposed methods.

Overall, the subjective evaluation demonstrates that the proposed numerical framework is responsible for a substantial portion of the perceptual improvement, while cross-attention feature injection provides complementary gains in editing performance. In particular, the dramatic performance gap between the original FluxMusic baseline and \textit{FlowSonic No Injection} highlights the effectiveness of the proposed Seeded-AB3 solver and DHC strategy in improving the stability and quality of zero-shot music editing. Building upon this stronger numerical foundation, the proposed feature injection mechanisms further improve semantic consistency and preservation of the original recording, with KV Injection providing the best overall balance and V Injection achieving the strongest preservation performance. Together, these components enable FlowSonic to consistently generate more natural, semantically accurate, and perceptually convincing music edits than existing zero-shot baseline methods.

\section{Conclusion}
\label{sec:conclusion}

In this work, we presented \textbf{FlowSonic}, a zero-shot framework for editing real-world music recordings using a pretrained rectified-flow diffusion transformer. Beyond introducing an effective editing framework, we investigated the role of numerical ODE solvers in inversion-based music editing and demonstrated that the choice of integration strategy has a substantial impact on trajectory stability, structural preservation, and semantic controllability. In particular, we showed that high-order Adams--Bashforth integration provides a more accurate approximation of the underlying rectified-flow dynamics, while its application to editing is limited by the absence of derivative history at the beginning of the generation process.

To address this limitation, we introduced a \emph{Seeded Adams--Bashforth} integration framework based on \emph{Dynamic History Caching}, which reuses velocity evaluations computed during deterministic inversion to initialize the multi-step solver. By eliminating the conventional lower-order warm-start stage, the proposed approach enables high-order integration from the first generation step while maintaining numerical consistency between inversion and generation. Extensive experiments on both timbre-transfer and genre-transformation tasks demonstrated that the proposed framework produces smoother latent trajectories and consistently improves semantic alignment, structural preservation, and perceptual audio quality compared with conventional solver configurations.

Although our framework is built upon FluxMusic, whose editing capability is inherently limited by the underlying pretrained model, the proposed numerical integration strategy consistently improves its editing performance without additional training or optimization. We believe that DHC is model-agnostic and can be readily incorporated into future rectified-flow generative models, providing a simple and efficient mechanism for improving inversion-based editing across a broad range of generative audio applications.

\bibliographystyle{IEEEtran}
\bibliography{references}

\clearpage
\appendix
\section{Appendix}

\subsection{Additional Results on Attention Injection Variants}
\label{sec:appendix}

\begin{figure}[h!]
  \centering
  \includegraphics[width=\linewidth]{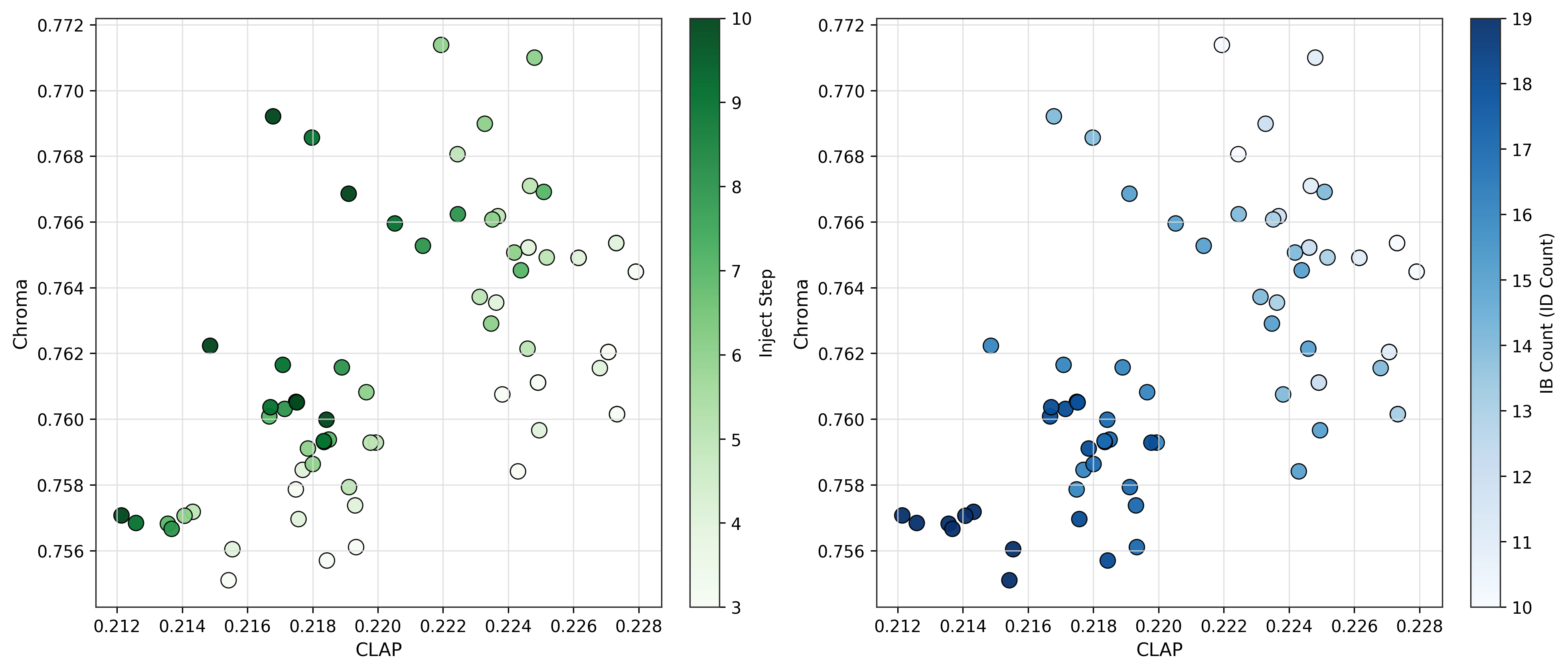}
  \caption{Results of injecting the key (\(K\)) components of the attention mechanism during timbre transfer task. Injecting \(K\) leads to moderate improvements in transferability but slightly weaker fidelity compared to \(V\)-injection, as less low-level acoustic information is preserved.}
  \label{fig:injection_K}
\end{figure}

\begin{figure}[h!]
  \centering
  \includegraphics[width=\linewidth]{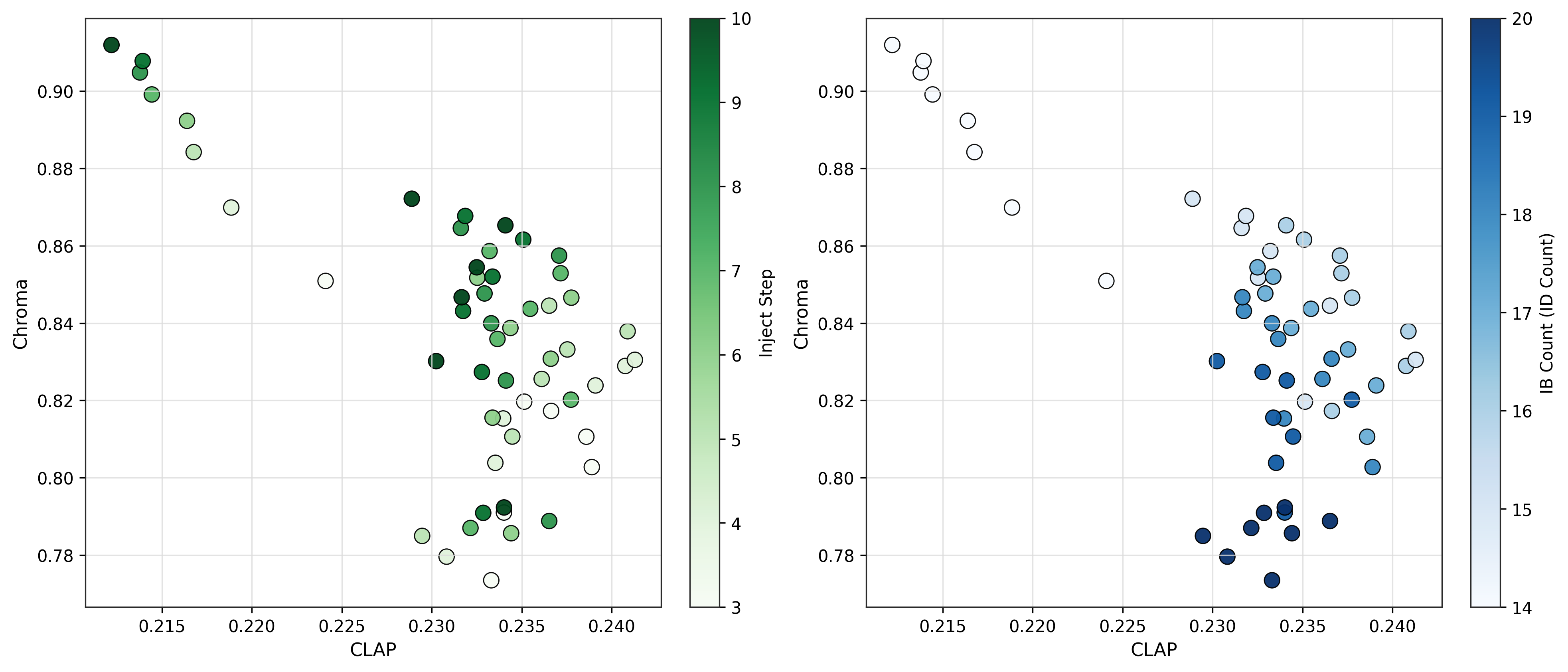}
  \caption{Results of injecting both key and value (\(K+V\)) components of the attention mechanism during timbre transfer task. Injecting \(K+V\) tends to balance fidelity and transferability, yielding more consistent timbre adaptation while retaining semantic control.}
  \label{fig:injection_KV}
\end{figure}

\begin{figure}[h!]
  \centering
  \includegraphics[width=\linewidth]{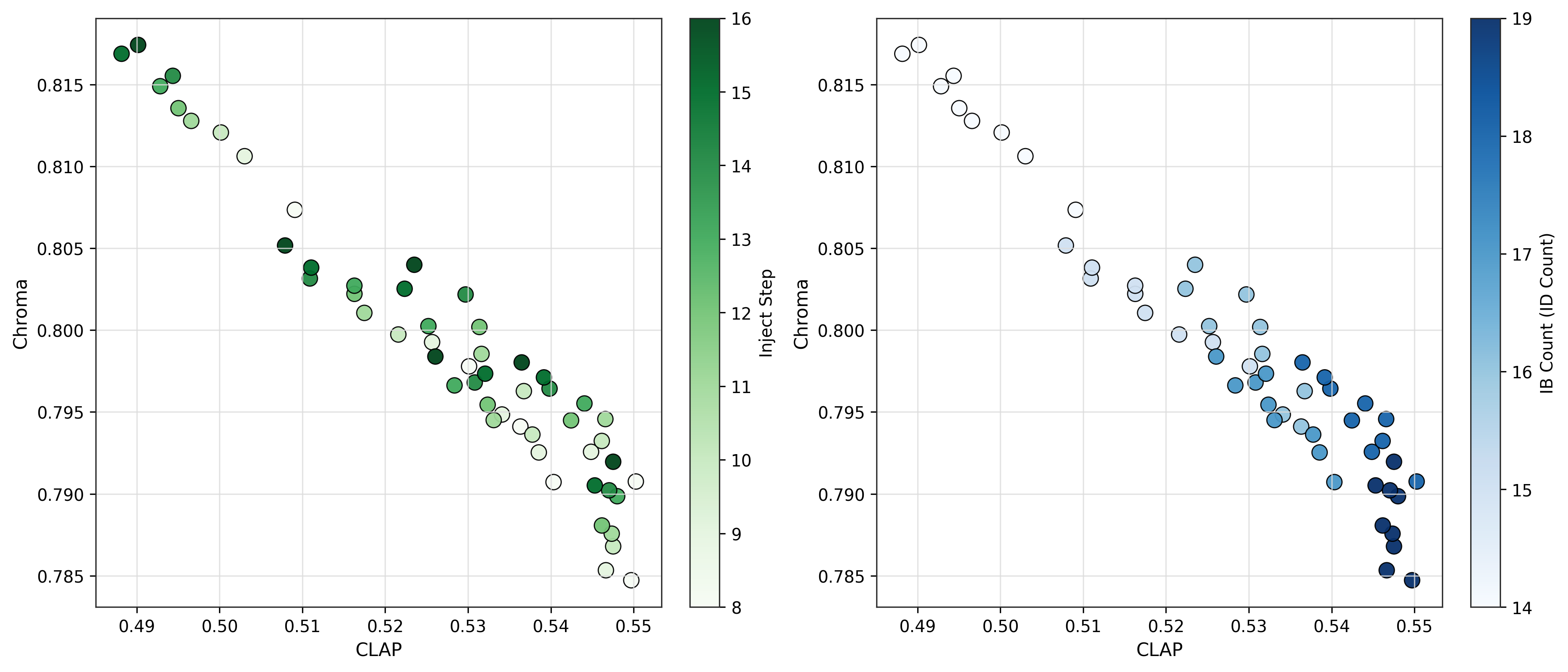}
  \caption{Results of injecting the value (V) components of the attention mechanism during genre transfer. Injecting V mainly preserves fidelity while limiting the degree of stylistic transfer, showing more stable tonal similarity across genres.}
  \label{fig:genre_injection_V}
\end{figure}

\begin{figure}[h!]
  \centering
  \includegraphics[width=\linewidth]{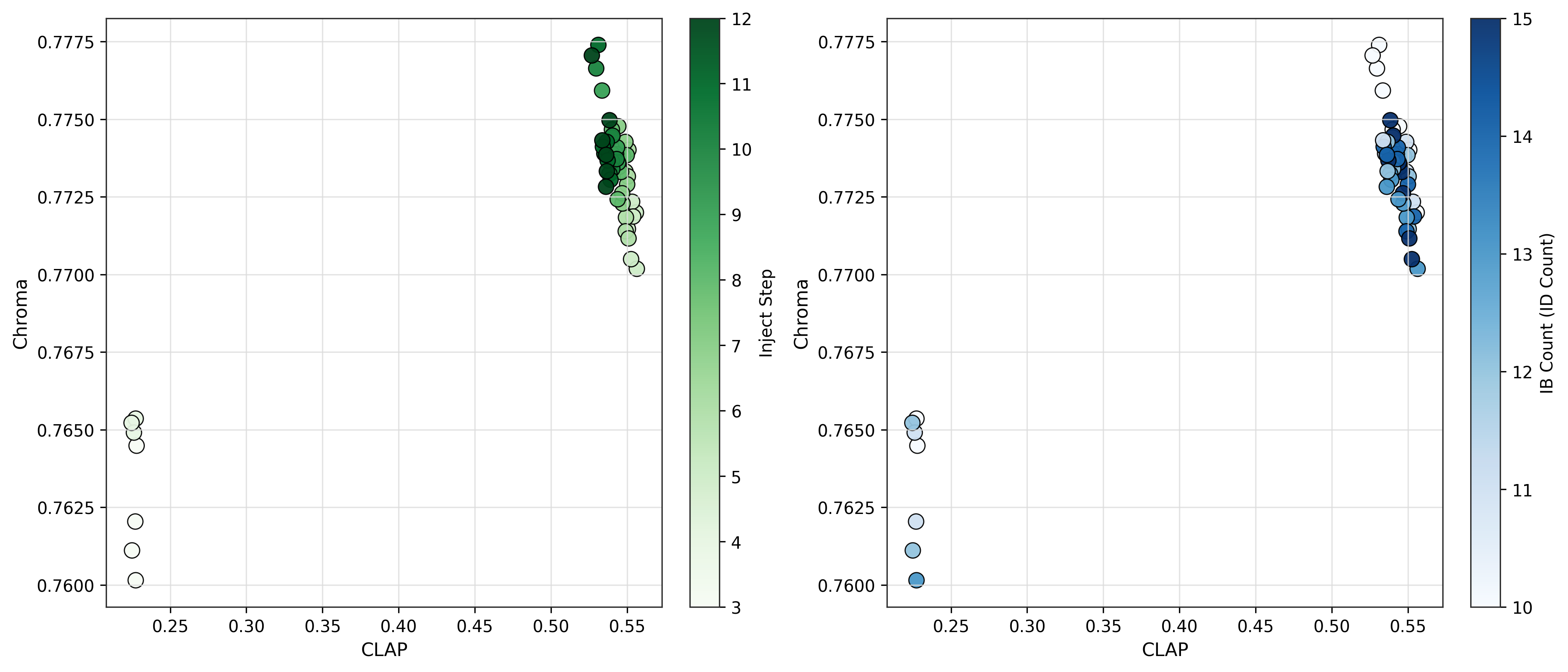}
  \caption{Results of injecting the key (K) components of the attention mechanism during genre transfer. Injecting only K emphasizes structural transferability but can reduce chroma fidelity, indicating that genre cues dominate over tonal preservation.}
  \label{fig:genre_injection_K}
\end{figure}

\begin{figure}[h!]
  \centering
  \includegraphics[width=\linewidth]{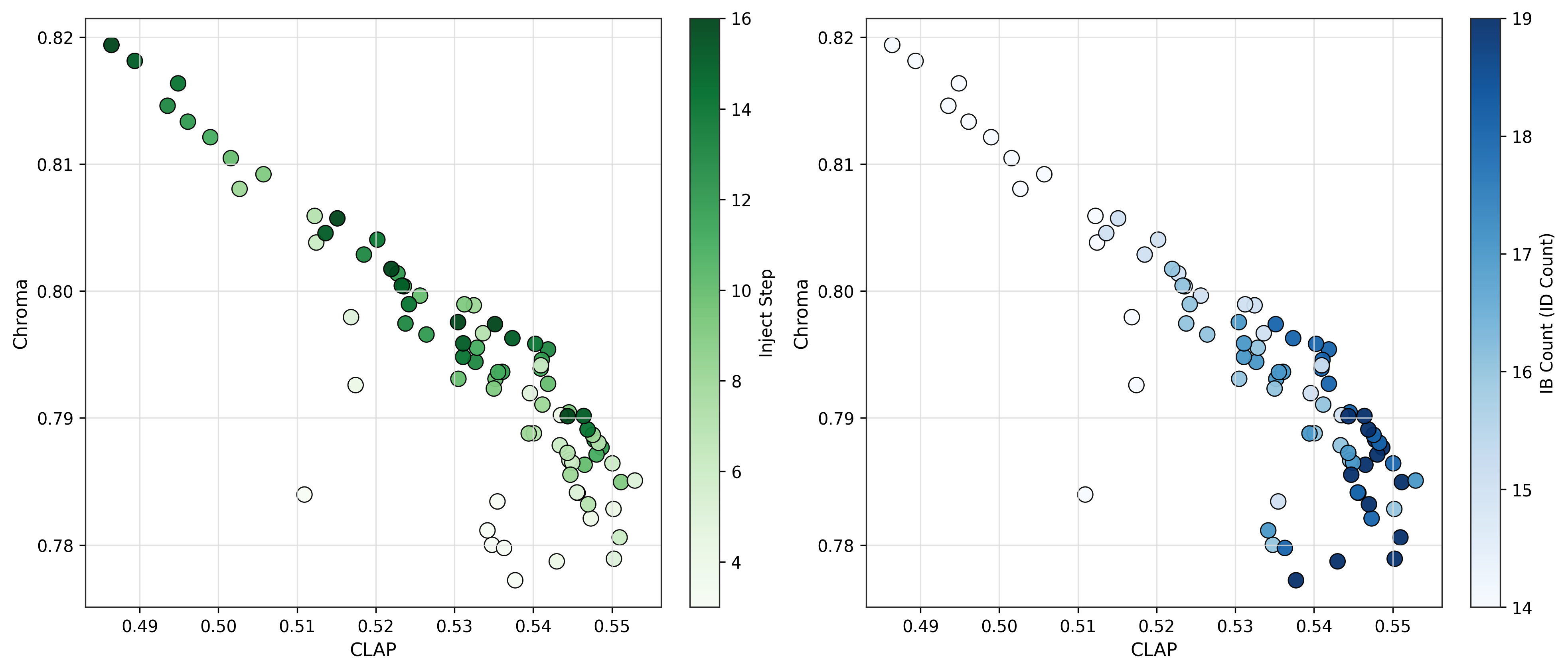}
  \caption{Results of injecting both key and value (K + V) components of the attention mechanism during genre transfer. Injecting K + V achieves a better balance between fidelity and transferability, enabling effective genre transformation while maintaining harmonic consistency.}
  \label{fig:genre_injection_KV}
\end{figure}

To complement the main results presented in Figure~\ref{fig:comparison}, we further examine the impact of injecting different attention components during generation, including the key (K) features and the combined key and value (K + V) features. In addition to the timbre-transfer experiments discussed in the main text, we also present the results obtained using value (V), key (K), and key + value (K + V) injection for the genre-transfer task. These additional experiments demonstrate that the choice of injected attention components significantly affects the trade-off between preservation of the original recording and adherence to the target editing prompt. Specifically, V injection better preserves timbral and genre-specific characteristics of the source audio, K injection encourages stronger conformity to the requested edit, while K + V injection provides a balanced trade-off, yielding consistent transformations that maintain both perceptual quality and structural coherence.

\subsection{Genre Trajectory Analysis}

\begin{figure*}[!t]
  \centering
  \includegraphics[width=0.9\textwidth]{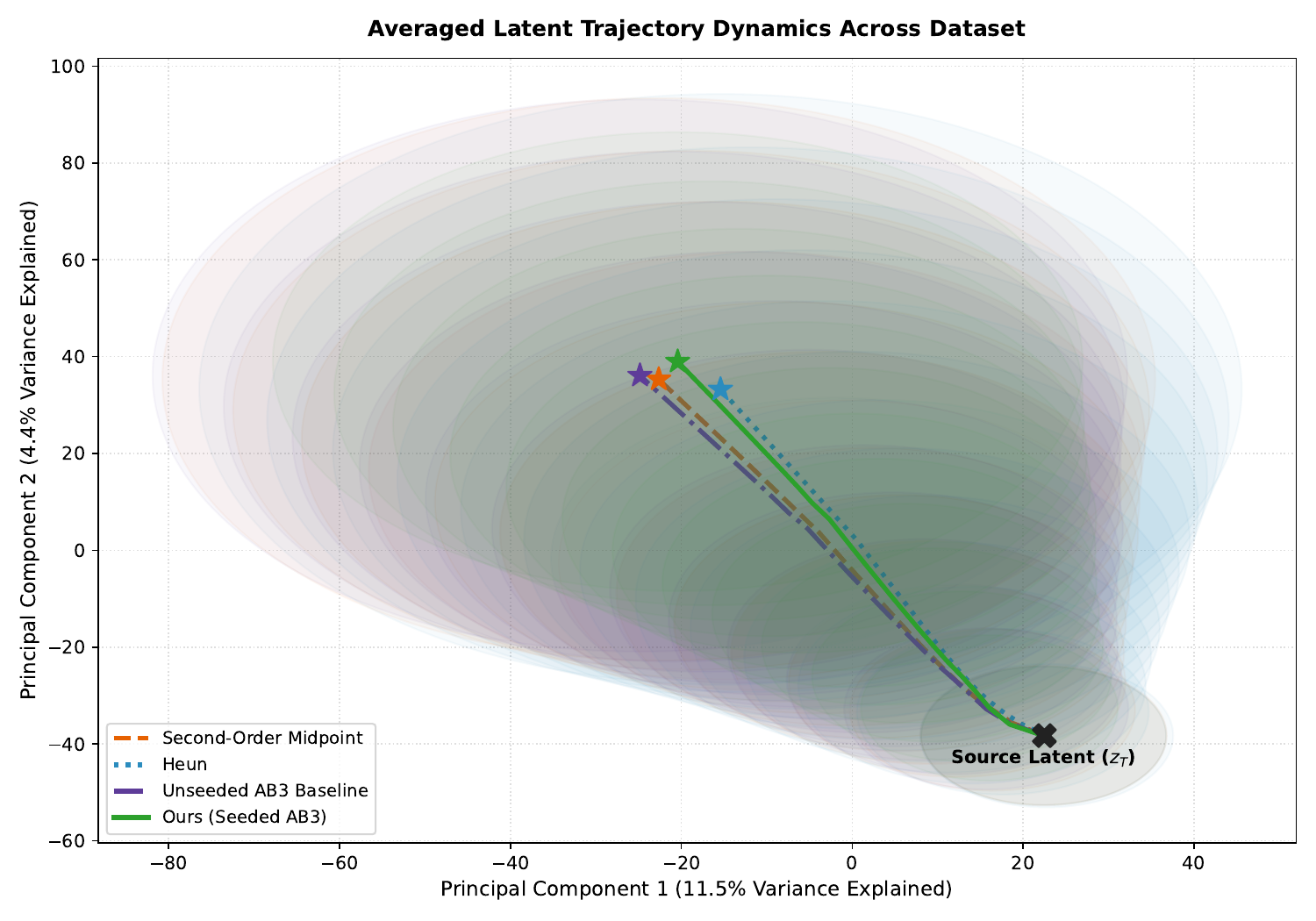}
  \caption{Global latent-space trajectory visualization for the genre-transfer task projected onto the first two principal components (PCA). The trajectories show the average evolution of latent states from the shared source latent ($\mathbf{z}_T$, black $\times$) to the final edited representations ($\star$), while the circles represent one standard deviation across the evaluation dataset at each integration step. Compared with the timbre-transfer task, all methods exhibit substantially broader trajectory dispersion, reflecting the larger latent-space displacement required for genre-level modifications. The Heun solver and both AB3-based methods converge toward similar target regions, whereas the Second-Order Midpoint solver follows a noticeably different transport direction. The proposed Seeded AB3 solver closely overlaps with the conventional AB3 trajectory, indicating that history-buffer initialization preserves the global transport path while improving the numerical behavior of the integration process.}
  \label{fig:trajectory_map_genre}
\end{figure*}

To investigate whether the Seeded-AB3 solver effects observed in timbre editing also extend to larger-scale transformations, we analyze the latent-space dynamics of different numerical solvers during genre transfer. Figure~\ref{fig:trajectory_map_genre} visualizes the average latent trajectories projected onto the first two principal components across the evaluation dataset.

Compared to the timbre-transformation setting, all solver trajectories exhibit substantially broader variance clouds, reflecting the increased structural complexity of genre modification. Because genre transfer affects multiple musical attributes simultaneously, including instrumentation, texture, and rhythmic characteristics, the resulting latent trajectories span a wider region of the projected latent space.

Despite this increased variability, the overall transport behavior remains remarkably consistent across most numerical solvers. The Heun solver and both AB3-based methods converge toward nearly the same region of the latent space, indicating that they learn a similar global editing direction despite the larger semantic changes required for genre transfer. In contrast, the Second-Order Midpoint solver terminates at a noticeably different location, suggesting that its numerical integration follows a different transport solution from the remaining methods.

Another important observation is that the trajectories of the proposed Seeded AB3 solver and the conventional AB3 solver almost completely overlap throughout the integration process. Unlike the timbre-transfer setting, where seeded initialization visibly altered the trajectory, genre transfer is dominated by a much larger global displacement, making the difference between the two AB3 variants less apparent in the two-dimensional PCA projection. Nevertheless, the close agreement between their trajectories indicates that the proposed history initialization preserves the desired transport direction rather than introducing unintended deviations.

Overall, the trajectory analysis indicates that the effect of the proposed Seeded AB3 initialization differs between timbre and genre editing. Unlike the timbre-transfer task, where Seeded AB3 produces a visibly different and more stable transport trajectory than the conventional AB3 solver, the trajectories in the genre-transfer task largely overlap. This is expected because genre editing requires substantially larger latent-space displacements, causing the global transport direction to dominate the two-dimensional PCA visualization and obscure finer differences between the two AB3 variants. Nevertheless, the quantitative results in Table~\ref{tab:solver_metrics} show that Seeded AB3 consistently achieves a better balance between semantic consistency and preservation of the original recording, indicating that its benefits arise from improved numerical stability during integration rather than a fundamentally different global transport path.

subsection{Full Subjective Results}

\begin{table*}[t]
\centering
\caption{The full subjective evaluation results on the timbre-transfer task.}
\begin{tabular}{lcccccc}
\toprule
 & \multicolumn{3}{c}{\textbf{MOS-T mean $\uparrow$}} & \multicolumn{3}{c}{\textbf{MOS-P mean $\uparrow$}} \\
\cmidrule(lr){2-4} \cmidrule(lr){5-7}
\textbf{Model} & \textbf{Overall} & \textbf{Professional Musicians} & \textbf{Ordinary Listeners} & \textbf{Overall} & \textbf{Professional Musicians} & \textbf{Ordinary Listeners} \\
\midrule
MusicGen & 3.75 & \textbf{4.30} & 3.20 & 2.20 & 2.30 & 2.10 \\
AudioLDM2 & 3.25 & 3.00 & 3.50 & 3.20 & 3.20 & 3.20 \\
Zeta & 3.25 & 3.30 & 3.20 & 3.45 & 3.30 & 3.60 \\
FluxMusic & 1.10 & 1.20 & 1.00 & 1.05 & 1.10 & 1.00 \\
\hline
FlowSonic No Injection (\textbf{ours}) & 3.55 & 3.50 & 3.60 & 3.35 & 3.50 & 3.20 \\
FlowSonic K Injection (\textbf{ours}) & 3.25 & 3.30 & 3.20 & 3.25 & 3.70 & 2.80 \\
FlowSonic KV Injection (\textbf{ours}) & \textbf{4.00} & \underline{4.10} & \underline{3.90} & \textbf{4.20} & \textbf{4.70} & 3.70 \\
FlowSonic V Injection (\textbf{ours}) & \textbf{4.00} & \underline{4.10} & \underline{3.90} & \underline{4.10} & \underline{4.30} & \textbf{3.90} \\
\bottomrule
\end{tabular}
\label{tab:full_subjective_timbre}
\end{table*}

\begin{table*}[t]
\centering
\caption{The full subjective evaluation results on the genre-transfer task.}
\begin{tabular}{lcccccc}
\toprule
 & \multicolumn{3}{c}{\textbf{MOS-T mean $\uparrow$}} & \multicolumn{3}{c}{\textbf{MOS-P mean $\uparrow$}} \\
\cmidrule(lr){2-4} \cmidrule(lr){5-7}
\textbf{Model} & \textbf{Overall} & \textbf{Professional Musicians} & \textbf{Ordinary Listeners} & \textbf{Overall} & \textbf{Professional Musicians} & \textbf{Ordinary Listeners} \\
\midrule
MusicGen & 2.45 & 2.80 & 2.10 & 2.05 & 2.00 & 2.10 \\
AudioLDM2 & \textbf{4.30} & \textbf{4.60} & \textbf{4.00} & 1.30 & 1.10 & 1.50 \\
Zeta & 3.00 & 3.40 & 2.60 & 3.10 & 3.20 & 3.00 \\
FluxMusic & 1.00 & 1.00 & 1.00 & 1.00 & 1.00 & 1.00 \\
\hline
FlowSonic No Injection (\textbf{ours}) & 3.50 & 3.40 & 3.60 & 2.95 & 3.40 & 2.50 \\
FlowSonic K Injection (\textbf{ours}) & 3.50 & 3.60 & 3.40 & 3.30 & 4.00 & 2.60 \\
FlowSonic KV Injection (\textbf{ours}) & \underline{3.95} & \underline{4.20} & \underline{3.70} & \underline{4.15} & \textbf{4.50} & 3.80 \\
FlowSonic V Injection (\textbf{ours}) & \textbf{4.00} & \underline{4.30} & \underline{3.70} & \textbf{4.35} & \textbf{4.50} & \textbf{4.20} \\
\bottomrule
\end{tabular}
\label{tab:full_subjective_genre}
\end{table*}

Tables~\ref{tab:full_subjective_timbre} and~\ref{tab:full_subjective_genre} present the complete subjective evaluation, including separate ratings from professional musicians and ordinary listeners. Although both groups exhibit similar overall preferences, they emphasize different aspects of the editing results. Professional musicians generally assign higher scores to models that accurately preserve the characteristics of the original recording while performing the requested edit, whereas ordinary listeners tend to favor outputs that sound more coherent and musically pleasing. Nevertheless, both groups consistently rank the proposed FlowSonic variants above the baseline methods, indicating that the improvements achieved by the proposed framework are robust across different levels of musical expertise.

A noteworthy observation is the substantial improvement obtained by introducing the proposed numerical framework alone. Even without cross-attention feature injection, \emph{FlowSonic No Injection} dramatically outperforms the original FluxMusic editing pipeline in both listener groups. For timbre transfer, the overall MOS-T increases from 1.10 to 3.55 and MOS-P from 1.05 to 3.35, while for genre transfer MOS-T improves from 1.00 to 3.50 and MOS-P from 1.00 to 2.95. Since no feature injection is employed in this variant, these gains originate entirely from the proposed Seeded-AB3 solver and DHC strategy, demonstrating that improved numerical integration alone substantially enhances semantic consistency with the editing prompt as well as preservation of the original recording.

For the timbre-transfer task, both listener groups strongly favor the proposed injected variants over all baseline methods. Professional musicians assign the highest MOS-P score to \emph{FlowSonic KV Injection} (4.70), indicating that combining key and value features most effectively preserves the musical characteristics of the source recording. Both KV and V Injection achieve identical MOS-T scores (4.10), suggesting comparable semantic consistency with the editing prompt. Ordinary listeners exhibit a similar ranking, but show a slight preference for \emph{FlowSonic V Injection}, which achieves the highest MOS-P score (3.90), while KV Injection follows closely (3.70). Among the baseline methods, MusicGen receives the highest MOS-T score from professionals (4.30), indicating strong semantic alignment, whereas Zeta achieves the best preservation score (MOS-P = 3.60) among ordinary listeners.

The genre-transfer task reveals similar trends. Professional musicians again favor the proposed FlowSonic variants, with both KV and V Injection achieving the highest MOS-P score (4.50), while V Injection attains the highest semantic consistency among the proposed methods (MOS-T = 4.30). Ordinary listeners also prefer the injected variants, assigning the highest MOS-P score to \emph{FlowSonic V Injection} (4.20), followed by KV Injection (3.80). Although AudioLDM2 achieves the highest MOS-T scores for both professional (4.60) and ordinary listeners (4.00), its low MOS-P scores (1.10 and 1.50, respectively) indicate that, despite closely following the target prompt, it fails to preserve the characteristics of the original recording. This contrast highlights the importance of simultaneously optimizing semantic consistency and source preservation for practical music editing.

Overall, the results demonstrate that the proposed framework benefits from two complementary components. The Seeded-AB3 integration strategy provides the primary improvement by substantially increasing editing stability and establishing a much stronger baseline than the original FluxMusic inference procedure. Building upon this foundation, cross-attention feature injection further improves both semantic consistency and preservation of the original recording. While professional musicians consistently prefer KV Injection because of its stronger preservation capability, ordinary listeners exhibit a slight preference for V Injection owing to its more balanced and perceptually coherent edits. The largely consistent ranking across both listener groups confirms that the proposed FlowSonic framework generalizes well across users with different levels of musical expertise.
\end{document}